# Syntax Repair as Language Intersection

BREANDAN CONSIDINE

Syntax repair can be viewed as a question about finite languages: given an invalid string and a grammar, which nearby strings are syntactically valid? We formalize bounded syntax repair as the intersection of a context-free language with an acyclic Levenshtein automaton, obtaining a finite language that contains all and only repairs within a chosen edit radius. This approach leverages the classic Bar-Hillel construction for CFL-regular intersection, and specializes it to acyclic automata, and yielding a highly parallel decision procedure for deciding repairability. We then show how Brzozowski derivatives support fast incremental enumeration from this intersection, separating exact admissibility from probabilistic ranking. Experiments on Python syntax errors demonstrate the resulting candidate spaces are practical to construct and substantially improve repair accuracy when used as a grammar-constrained search space.

## 1 INTRODUCTION

A missing delimiter or mistyped keyword can turn an otherwise meaningful program into an unrecognizable string. Typical parsers can detect invalidity, but cannot infer what the author intended to write. Suppose the parser instead responds by proposing a small set of valid alternatives. What should that set contain? We propose that each suggestion be (1) syntactically valid, (2) similar to the original string, (3) highly natural. Producing a single repair with any one of these properties individually is relatively straightforward. Producing a small set of repairs satisfying all three properties – without missing the author's intended repair – is substantially harder.

This paper studies that problem for syntax defined by an arbitrary context-free grammar. Let $G$ describe the language, let $\underset{\sim}{\sigma} \notin \mathcal{L}(G)$ be an invalid token sequence, and let $L(\underset{\sim}{\sigma}, d)$ denote the strings within $d$ edits of $\underset{\sim}{\sigma}$. The natural candidate space is $\ell_\cap = \mathcal{L}(G) \cap \mathcal{L}\big(L(\underset{\sim}{\sigma}, d)\big)$. The grammar enforces validity, the edit language enforces locality, and a probabilistic model supplies naturalness. Assuming the intended repair requires at most $d$ edits, it will be contained in $\ell_\cap$ and the problem becomes how to represent this finite language and quickly recover its members.

Classical error-correcting parsers return either one repair or a set minimizing a hand-designed cost [2, 20]. While appealing for their interpretability and well-understood algorithmic properties, these methods are unable to handle ambiguity and must rely on relatively brittle heuristics to select the repair. Learning-based systems [54, 66], while statistically more robust, typically use some form of approximate inference and thus require expensive postprocessing or rejection sampling to ensure the generated results conform to the grammar. In both cases, *existing approaches generate far too few repairs to be effective*. We argue that for local syntax errors, completeness must be a first-class objective: a repair system must retrieve broadly before selecting the best repair(s).

Our approach makes the local search space explicit. We construct an acyclic Levenshtein automaton denoting every string within distance $d$ of $\underset{\sim}{\sigma}$, then intersect it with the grammar. Previously this route may have deterred exploration to its seemingly high cost: directly materializing the Bar–Hillel product appears intractable for practical grammars. Instead, we exploit a correspondence between Bar–Hillel intersection and CFL reachability to quickly construct the intersection grammar.

This representation turns repair into a constrained decoding problem, allowing any off-the-shelf autoregressive decoder to traverse $\ell_\cap$ incrementally without rejection sampling. In our implementation, a lightweight decoder provides an initial set of plausible hypotheses, of which the top-k are then scored with a pretrained transformer, reranked, and finally truncated. Syntactic validity is enforced by construction, edit locality becomes exact, and transformer-based reranking is reserved for the highest-probability results, ensuring low latency and high naturalness.

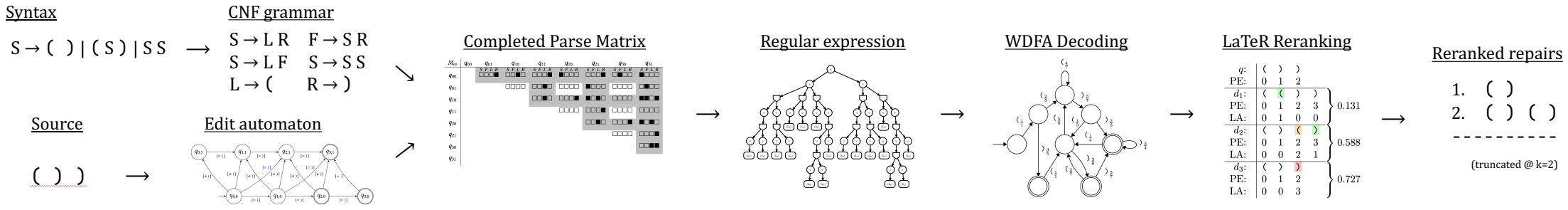

Fig. 1. We first represent the bounded edit neighborhood as an automaton, intersect it with the language grammar, and finally construct a regular expression representing the intersection between the language and edit neighborhood. A fast probabilistic decoder enumerates probable members from this intersection, and a learned language model reranks the highest-probability candidates for presentation.

This paper makes three contributions. First, it formulates bounded syntax repair as exact language intersection and gives a practical construction for arbitrary context-free grammars. Second, it gives a parallel matrix-closure procedure for intersection nonemptiness of a CFG and an acyclic NFA, proving that the closure phase cannot require more than a linear number of squaring rounds, and compares various parallel decoding algorithms. Third, it implements and evaluates this entire repair pipeline on multicore CPUs and GPUs. On naturally occurring Python errors of up to three lexical edits and eighty tokens, this tool operates within subsecond latency and achieves up to a fivefold improvement in top-1 repair precision over prior systems under comparable resource limits.

## 2 BACKGROUND

Recall that a CFG, $\mathcal{G} = \langle \Sigma, V, P, S \rangle$, is a quadruple consisting of terminals ($\Sigma$), nonterminals ($V$), productions $\big(P \colon V \to (V \mid \Sigma)^+\big)$, and a start symbol, ($S$). Every CFG is reducible to so-called *Chomsky Normal Form* [17], $P' \colon V \to (V^2 \mid \Sigma)$, where every production is either (1) a binary production $w \to xz$, or (2) a terminal production $w \to t$, where $w, x, z : V$ and $t : \Sigma$. For example:

$$G = \big\{\, S \to S\,S \mid (\,S\,) \mid (\,) \,\big\} \Longrightarrow G' = \big\{\, S \to L\,R \mid L\,F \mid S\,S,\ F \to S\,R,\ L \to (,\ R \to ) \,\big\}$$

Likewise, a finite automaton (FA) is a quintuple $\mathcal{A} = \langle Q, \Sigma, \delta, q_\alpha, F \rangle$, where $Q$ is a finite set of states, $\Sigma$ is a finite alphabet, $\delta \subseteq Q \times \Sigma \times Q$ is the transition relation, $q_\alpha$ is the initial state, and $F \subseteq Q$ are the accepting states. These generally come in two varieties, deterministic and nondeterministic, depending on whether $\delta$ maps each pair $\langle q, s \rangle$ to at most one $q'$. A weighted finite automaton (WFA) generalizes the FA over a weighted semiring, $\mathbb{S}$, with a transition relation $\delta \colon Q \times \Sigma \times Q \to \mathbb{S}$ and boundary weights $I, F \colon Q \to \mathbb{S}$. Primarily, we will work over $\mathbb{B}$ in the unweighted setting and $\mathbb{R}_+$ in the weighted setting. We write $\mathcal{L}(\cdot)$ for the language generated by an automaton or grammar.

Although the family of context-free languages is not closed under self-intersection, it has the useful property of being closed under intersection with regular languages [7]:

Theorem 2.1 (Bar-Hillel, 1961). *For any CFG, $G = \langle \Sigma, V, P, S \rangle$, and nondeterministic finite automaton (NFA), $A = \langle Q, \Sigma, \delta, q_\alpha, F \rangle$, there is a CFG, $G_\cap = \langle \Sigma_\cap, V_\cap, P_\cap, S_\cap \rangle$ s.t. $\mathcal{L}(G_\cap) = \mathcal{L}(G) \cap \mathcal{L}(A)$.*

Salomaa [55] gives the following explicit, but inefficient construction for $G_\cap$:

*Definition 2.2 (Salomaa, 1973).* Assume $G$ is in CNF. For every nonterminal $w \in V$ and state pair $p, r : Q$, there will be a synthetic nonterminal $pwr$ in $V_\cap$. Now, for every production $(w \to xz) \in P'$ and state triple $p, q, r : Q$, add $pwr \to (pxq)(qzr)$ to $P_\cap$; for every production $(w \to a) \in P'$ and transition $q \overset{a}{\to} r$, add $qwr \to a$ to $P_\cap$. Lastly, for every final state $q_\omega$, add $S_\cap \to q_\alpha S q_\omega$ to $P_\cap$.

This construction will generate a large number of useless productions containing state pairs that are not even connected by any path, which is clearly excessive. Furthermore, most synthetic productions in $P_\cap$ will be non-generating or unreachable, requiring subsequent normalization to restore CNF. In § 3, we will present a far more efficient construction for the special case when the intersection is finite. But first, let us return to the broader question of syntax repair.

## 2.1 Informal statement

Assume there exists a transducer from raw text to grammatical tokens, $t : \Sigma_U^* \to \Sigma_G^*$. In the compiler nomenclature, $t$ is called a *lexer* and would typically be regular under mild conditions. In this paper, we do not consider $t$ and strictly deal with languages over $\Sigma_G^*$, or simply $\Sigma^*$ for brevity.

Now suppose we have a syntax, $\ell \subset \Sigma^*$, containing every acceptable program. A syntax error is an unacceptable string, $\underset{\sim}{\sigma} \notin \ell$, that we wish to repair. We can model syntax repair as a language intersection between a context-free language (CFL) and a regular language. Henceforth, $\underset{\sim}{\sigma}$ will always and only be used to denote a syntactically invalid string whose intended language is known.

Given a lexical representation of a broken computer program $\underset{\sim}{\sigma}$ and a grammar $G$, our goal is to find every valid string $\sigma$ consistent with the grammar $G$ and within a certain edit distance, $d$. Consider the language of nearby strings: if intersected with the language of grammatically valid programs, $\mathcal{L}(G)$, the result ($\ell_\cap$) will contain every possible repair within the given edit distance, a subset of which will be natural or statistically probable. We will sample these repairs, map them back into Unicode, adding placeholders for fresh names, numbers, and string literals, then finally apply an off-the-shelf code formatter to display them. Both the preprocessing and the cosmetic postprocessing steps are ancillary to the scope of this work, in which we confine ourselves to a lexical alphabet.

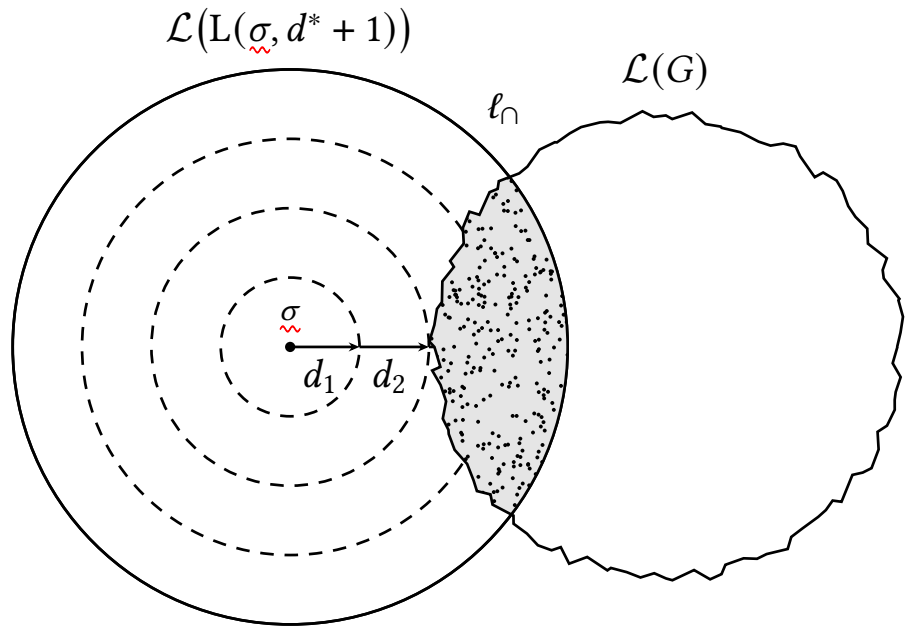

Fig. 2. CFL intersection with the local edit region around a broken code snippet, where $d^* = 3$ is the language edit distance (LED).

## 2.2 Formal statement

Let us now restate our informal description of the syntax repair problem in more formal terms.

*Definition 2.3 (Bounded Levenshtein-CFL reachability).* Given a CFL, $\ell$, and an invalid string, $\underset{\sim}{\sigma} : \bar{\ell}$, find every valid string reachable within $d$ edits of $\underset{\sim}{\sigma}$, i.e., letting $\Delta$ be the Levenshtein metric and $\mathcal{L}\big(L(\underset{\sim}{\sigma}, d)\big) = \{\sigma' \mid \Delta(\underset{\sim}{\sigma}, \sigma') \leq d\}$ be the Levenshtein $d$-ball, we seek to find $\ell_\cap = \mathcal{L}\big(L(\underset{\sim}{\sigma}, d)\big) \cap \ell$.

As the intersection language, $\ell_\cap$, typically contains a large number of possible repairs, we want a procedure that surfaces both natural and valid repairs over unnatural but valid repairs:

*Definition 2.4 (Ranked repair).* Given a finite language $\ell_\cap = \mathcal{L}\big(L(\underset{\sim}{\sigma}, d)\big) \cap \ell$ and a probabilistic language model $\mathrm{P}_\theta : \Sigma^* \times \underset{\sim}{\Sigma}^* \to \mathbb{R}$, find the top-$k$ highest probability repairs. That is,

$$R(\ell_\cap, P_\theta) : 2^{\Sigma^*} \times (\Sigma^* \to \mathbb{R}) \to (\Sigma^*)^{\leq k} = \underset{\boldsymbol{\sigma} \subseteq \ell_\cap, |\boldsymbol{\sigma}| \leq k}{\operatorname{argmax}} \sum_{\sigma \in \boldsymbol{\sigma}} \mathrm{P}_\theta(\sigma \mid \underset{\sim}{\sigma}) \tag{1}$$

A popular approach to ranked repair involves learning a distribution over strings, however this is highly sample-inefficient and generalizes poorly to new languages. Approximating a distribution over $\Sigma^*$ forces the model to jointly learn syntax and stylometry. Furthermore, even with an extremely efficient approximate sampler for $\sigma \leftsquigarrow \ell_\cap$, due to the size of the languages involved, it would be intractable to sample either $\ell$ or $\mathcal{L}\big(L(\underset{\sim}{\sigma}, d)\big)$, reject duplicates, then reject unreachable or invalid edits, and completely out of the question to sample $\sigma \leftsquigarrow \Sigma^*$, as do most neural language models.

As we will demonstrate, the ranked repair problem can be factorized into three subproblems: (1) exact representation, (2) decoding and (3) reranking. Instead of working with strings, we will explicitly construct a grammar which soundly and completely generates $\ell \cap \mathcal{L}\big(L(\underset{\sim}{\sigma}, d)\big)$, then decode repairs from its language. By designing a sufficiently precise and extensive decoder, ensuring the retrieved set contains the true repair can be achieved with a much simpler, syntax-oblivious model. Finally, we will train a language model to rerank the repair candidates and take the top-$k$ results.

## 3 METHOD

Our key observation is that repairability can be treated as an instance of CFL reachability. Given a CFG $G$ and an acyclic NFA $A_\varnothing$, we compute which nonterminals can derive the paths $p \rightsquigarrow r$ in $A_\varnothing$, then ask whether some accepting path from $q_\alpha$ derives $S$. As long as one of the participants in the language intersection is presented as an acyclic FSA, we show the intersection nonemptiness problem is decidable in linear parallel time in the span of the transition relation. Formally,

THEOREM 3.1. *Let* $G = \langle \Sigma, V, P, S \rangle$ *be a CFG in CNF and let* $A_\varnothing = \langle Q, \Sigma, \delta, q_\alpha, F \rangle$ *be an* $\varepsilon$*-free acyclic NFA (ANFA). Define its span as* $|A_\varnothing| = \max\{m \mid A_\varnothing \text{ has an } m\text{-arc path}\}$. *There exists a decision procedure* $\Psi : \mathit{CFG} \to \mathit{ANFA} \to \mathbb{B}$ *such that* $\Psi(G, A_\varnothing) \vDash [\mathcal{L}(G) \cap \mathcal{L}(A_\varnothing) \neq \varnothing]$ *that requires* $\mathcal{O}(|A_\varnothing|)$ *time using* $\mathcal{O}(|Q|^3|P|)$ *parallel processors.*

PROOF. To prove nonemptiness, we must show there exists a path $q_\alpha \rightsquigarrow q_\omega$ in $A_\varnothing$ s.t. $q_\omega \in F$ where $q_\alpha \rightsquigarrow q_\omega \vdash S$. At least one of two cases must hold for $w \in V$ to parse a given $p \rightsquigarrow r$ pair:

(1) $p$ steps directly to $r$ in which case it suffices to check $\exists a.\big((p \xrightarrow{a} r) \in \delta \land (w \to a) \in P\big)$, or,

(2) there is some midpoint $q \in Q$, $p \rightsquigarrow q \rightsquigarrow r$ such that $\exists x, z.\big((w \to xz) \in P \land \overbrace{\underbrace{p \rightsquigarrow q}_{x}, \underbrace{q \rightsquigarrow r}_{z}}^{w}\big)$.

Since we assumed $\delta$ is acyclic, there exists a topological sort of $\delta$ imposing a total order on $Q$ such that the chart is strictly upper triangular (SUT). Note $Q$ can be ordered topologically in $\mathcal{O}(\log^2 |Q|)$ time [19] using matrix multiplication. Let $\mathcal{C} = \mathbb{B}^{|Q|\times|Q|\times|V|}$ be Boolean charts with pointwise join $\vee$ and bottom $\bot$. Define the product

$$(X \otimes Y)[p, r, w] = \bigvee_{q \in Q,\ x, z \in V} X[p, q, x] \land Y[q, r, z] \land [(w \to xz) \in P]$$

and let addition be logical disjunction. We call $(\mathcal{C}, \vee, \otimes, \bot)$ the chart semiring. Initialize

$$M[p, r, w] = \bigvee_{a \in \Sigma} \big[(p \xrightarrow{a} r) \in \delta \land (w \to a) \in P\big].$$

Borrowing the exponentiation-by-squaring[1] notation, set

$$T_0 = M, \qquad T_{j+1} = T_j \vee (T_j \otimes T_j), \qquad \exp(M) = T_\infty$$

Let $H_j[p, r, w]$ assert that some word labeling a path $p \rightsquigarrow r$ is derived from $w$ by a CNF parse tree of height at most $j$, with terminal productions at height 0. We prove $T_j[p, r, w] \Leftrightarrow H_j[p, r, w]$ by induction on $j$. The base case is exactly $M$. For the inductive case, a height-$(j+1)$ parse is either already of height at most $j$, or has root $w \to xz$ and a pivot state $q$ whose two subpaths have height-$\leq j$ parses from $x$ and $z$; by the induction hypothesis, this is exactly $T_{j+1}$.

Since $A_\varnothing$ is $\varepsilon$-free, a path of $m$ arcs has a label of length $m$. Every CNF parse of such a label has at most $m$ terminal leaves and height $< m$, so no productive chart fact over $A_\varnothing$ can require height exceeding $|A_\varnothing|$. Hence $T_{|A_\varnothing|}$ is the least chart containing $M$ and closed under $\otimes$, in particular $T_{|A_\varnothing|} = T_{|A_\varnothing|+1}$. Finally,

$$\Psi(G, A_\varnothing) = \bigvee_{q_\omega \in F} \exp(M)[q_\alpha, q_\omega, S],$$

which is true exactly when some accepting path label lies in $\mathcal{L}(G)$, equivalently when $\mathcal{L}(G) \cap \mathcal{L}(A_\varnothing) \neq \varnothing$. The test may be checked after each round, giving positive early termination. □

[1] By slight abuse of notation: customarily, semiring multiplication is associative and unital; here $\otimes$ is merely monotone and join-distributive, so exp denotes the exact parenthesized iteration. See Bernardy and Jansson [8] for further details.

Theorem 3.2 (Witness construction). *Under the hypotheses of Theorem 3.1, the saturated chart can be lifted to a finite star-free regular expression (RE),* $e_\cap$ *with* $\mathcal{L}(e_\cap) = \mathcal{L}(G) \cap \mathcal{L}(A_\varnothing)$. *Thus,* $\Psi(G, A_\varnothing)$ *implies a concrete string witnessing intersection nonemptiness exists.*

Proof. We will work over an algebra of $(\varepsilon, \wedge, \text{star})$-free REs plus $\varnothing$, $E_+ \ni e ::= \varnothing \mid \Sigma \mid e \vee e \mid e \cdot e$. Let us define a provenance semiring variant of Thm. 3.1, using $E_+$ union and concatenation:

$$R_0[p, r, w] = \bigvee_{\substack{a \in \Sigma \\ (p \overset{a}{\to} r) \in \delta \\ (w \to a) \in P}} a, \qquad [X \odot Y][p, r, w] = \bigvee_{\substack{q \in Q,\, x, z \in V \\ (w \to xz) \in P}} X[p, q, x] \cdot Y[q, r, z], \qquad R_{j+1} = R_j \vee (R_j \odot R_j),$$

with empty joins equal to $\varnothing$. For parse height $\leq j$, the invariant is

$$\mathcal{L}\big(R_j[p, r, w]\big) = \{\sigma \mid p \overset{\sigma}{\rightsquigarrow} r \text{ in } A \wedge w \Rightarrow_{\leq j} \sigma\}. \tag{*}$$

Similar to Thm. 3.1: terminals prove the base case, and each step chooses a rule $w \to xz$ and midpoint $q$. Acyclicity bounds accepting paths, hence the bounded closure reaches $R_\infty$. Taking $e_\cap = \bigvee_{q_\omega \in F} R_\infty[q_\alpha, q_\omega, S]$ (*), gives the claimed language equality. Since products use strict subpaths in topological order, the shared representation is an acyclic RE DAG. If nonempty, pruning $\varnothing$ yields a productive $E_+$, to which Lemma 3.4 applies. □

*Definition 3.3 (Brzozowski derivative).* Following Brzozowski [11], let $E$ be the star-free RE, $E \ni e ::= \varnothing \mid \varepsilon \mid \Sigma \mid e \vee e \mid e \wedge e \mid e \cdot e$. Write $\delta(e) \in \{\varnothing, \varepsilon\}$ for the nullable projection, where $\delta(e) = \varepsilon$ iff $\varepsilon \in \mathcal{L}(e)$. For $a, b \in \Sigma$, define a quotient operator $\partial_a : E \to E$:

$$\begin{aligned}
\partial_a(\ \varnothing\ ) &= \varnothing & \qquad\qquad \delta(\ \varnothing\ ) &= \varnothing \\
\partial_a(\ \varepsilon\ ) &= \varnothing & \delta(\ \varepsilon\ ) &= \varepsilon \\
\partial_a(\ b\ ) &= \begin{cases} \varepsilon & \text{if } a = b \\ \varnothing & \text{if } a \neq b \end{cases} & \delta(\ a\ ) &= \varnothing \\
\partial_a(\ x \cdot z\ ) &= (\partial_a x) \cdot z \vee \delta(x) \cdot \partial_a z & \delta(\ x \cdot z\ ) &= \delta(x) \wedge \delta(z) \\
\partial_a(x \vee z) &= \partial_a x \vee \partial_a z & \delta(x \vee z) &= \delta(x) \vee \delta(z) \\
\partial_a(x \wedge z) &= \partial_a x \wedge \partial_a z & \delta(x \wedge z) &= \delta(x) \wedge \delta(z)
\end{aligned}$$

Extend to strings by $\partial_\varepsilon e = e$ and $\partial_{a\sigma} e = \partial_\sigma(\partial_a e)$; then $\mathcal{L}(\partial_\sigma e) = \{\tau \mid \sigma\tau \in \mathcal{L}(e)\}$.

Lemma 3.4 (Generation from $E_+$). *Let* $E_+ \ni e ::= \Sigma \mid e \cdot e \mid e \vee e$ *be productive, i.e., every subexpression has nonempty language. With derivatives simplified by the identities for* $\varepsilon$ *and* $\varnothing$, *define*

$$\texttt{next}(e) = \begin{cases} \{a\} & e = a \in \Sigma, \\ \texttt{next}(x) & e = x \cdot z, \\ \texttt{next}(x) \cup \texttt{next}(z) & e = x \vee z, \end{cases}$$

$$\texttt{choose}(e) = \begin{cases} a & e = a \in \Sigma, \\ \big(s \leftsquigarrow \texttt{next}(e)\big) \cdot \texttt{choose}(\partial_s e) & e = x \cdot z, \\ \texttt{choose}\big(e' \leftsquigarrow \{x, z\}\big) & e = x \vee z, \end{cases}$$

*where* $\leftsquigarrow$ *denotes random choice from a set. Then* $\texttt{choose}(e)$ *terminates and returns some* $\sigma \in \mathcal{L}(e)$.

Proof. Structurally, $\texttt{next}(e) = \{a \in \Sigma \mid \exists \tau.\ a\tau \in \mathcal{L}(e)\}$, so every selected branch or derivative residual is nonempty. A concatenation step emits one symbol and recurses on a residual with shorter words; a union step descends to a strict subexpression. Finiteness rules out infinite descent. If the emitted word is $\sigma$, then $\varepsilon \in \mathcal{L}(\partial_\sigma e)$, hence $\sigma \in \mathcal{L}(e)$. □

## 4 EXAMPLES

In this section, we will consider three examples of intersections with finite languages. First, parsing can be viewed as a special case of intersection with a singleton language. Second, we will introduce *completion* as an intersection that admits terminal wildcards in fixed locations. Third, we consider syntax repair, where we will intersect a language representing all possible edit paths within a certain distance to determine the location(s) and fill them with the appropriate terminal(s).

### 4.1 Recognition as intersection

In the case of ordinary CFL recognition, the automaton forms a single row and accepts one word:

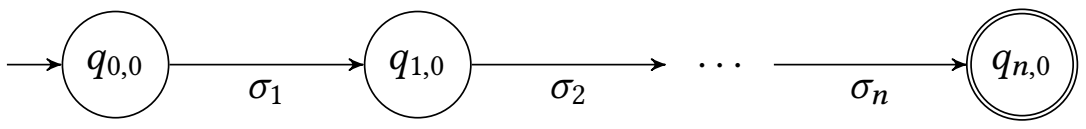

Since the word is predetermined, we just need to keep track of nonterminal subsets for each substring. So, given a CFG, $G' : \mathcal{G}$ in Chomsky Normal Form (CNF), we can construct a recognizer for strings $\sigma : \Sigma^n$ as follows. Let $2^V$ be our domain, 0 be $\varnothing$, $\oplus$ be $\cup$, and $\otimes$ be defined as:

$$X \otimes Z = \left\{ \; w \mid \langle x, z \rangle \in X \times Z, (w \rightarrow xz) \in P \; \right\} \tag{2}$$

If we define $\hat{\sigma}_r = \{w \mid (w \rightarrow \sigma_r) \in P\}$, then construct a matrix with nonterminals on the superdiagonal representing each token, $M_0[r + 1 = c](G', \sigma) = \hat{\sigma}_r$, the fixpoint $M_{i+1} = M_i + M_i^2$ is uniquely determined by the superdiagonal entries. Omitting the exponentiation-by-squaring detail, the ordinary fixedpoint iteration simply fills successive diagonals:

$$M_0 = \begin{pmatrix} \varnothing & \hat{\sigma}_1 & \varnothing & \cdots & \varnothing \\ \vdots & & \ddots & \ddots & \vdots \\ \vdots & & & \ddots & \varnothing \\ \vdots & & & \ddots & \hat{\sigma}_n \\ \varnothing & \cdots & \cdots & \cdots & \varnothing \end{pmatrix}, M_1 = \begin{pmatrix} \varnothing & \hat{\sigma}_1 & \Lambda & \cdots & \varnothing \\ \vdots & & \ddots & \ddots & \vdots \\ \vdots & & & \ddots & \Lambda \\ \vdots & & & \ddots & \hat{\sigma}_n \\ \varnothing & \cdots & \cdots & \cdots & \varnothing \end{pmatrix}, \ldots \quad , M_\infty = \begin{pmatrix} \varnothing & \hat{\sigma}_1 & \Lambda & \cdots & \Lambda_\sigma^* \\ \vdots & & \ddots & \ddots & \vdots \\ \vdots & & & \ddots & \Lambda \\ \vdots & & & \ddots & \hat{\sigma}_n \\ \varnothing & \cdots & \cdots & \cdots & \varnothing \end{pmatrix} \tag{3}$$

Once the fixpoint $M_\infty$ is attained, the proposition $[S \in \Lambda_\sigma^*]$ [2] decides language membership, i.e., $[\sigma \in \mathcal{L}(G)]$. So far, this procedure is essentially the textbook CYK algorithm in a linear algebraic notation [28] and a well-established technique in the parsing literature [29].

### 4.2 Completion as intersection

Let us now consider a problem of intermediate difficulty, wherein we are given a string template admitting edits at fixed locations, which can be filled by any terminal. When intersected with a CFL, this specifies a finite language whose contents are the set of all words consistent with the template. We call this problem *completion*. Formally,

*Definition 4.1 (Completion).* Let $\underline{\Sigma} = \Sigma \cup \{_\}$, where _ denotes a hole. We denote $\sqsubseteq : \Sigma^n \times \underline{\Sigma}^n$ as the relation $\{\langle \sigma', \sigma \rangle \mid \sigma_i \in \Sigma \implies \sigma_i' = \sigma_i\}$ and the set of all members $\{\sigma' : \Sigma^n \mid \sigma' \sqsubseteq \sigma\}$ as $\mathrm{H}(\sigma)$. Given a *porous string*, $\sigma : \underline{\Sigma}^n$ we seek all syntactically valid members, i.e., $A(\sigma) = \mathrm{H}(\sigma) \cap \ell$.

Here, the FSA takes a similar shape but can have multiple arcs between adjacent states, e.g.:

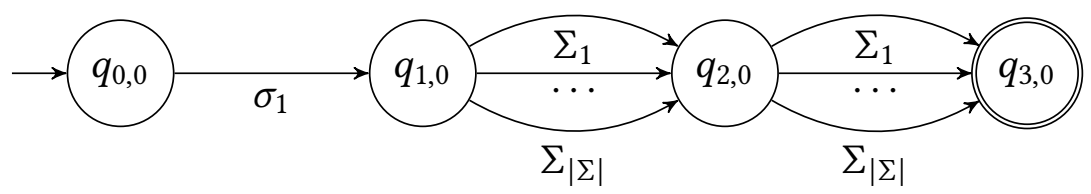

[2] Hereinafter, we use Iverson brackets to denote the indicator function of a predicate with free variables, i.e., $[P] \Leftrightarrow \mathbb{1}[P]$.

This corresponds to a template with two holes, $\sigma = 1$ _ _. Suppose the context-free grammar is $G = \{S \rightarrow NON, O \rightarrow + \mid \times, N \rightarrow 0 \mid 1\}$. This grammar will first be rewritten into CNF as $G' = \{S \rightarrow NL, N \rightarrow 0 \mid 1, O \rightarrow \times \mid +, L \rightarrow ON\}$. Using the powerset algebra we just defined, the matrix fixpoint $M' = M + M^2$ can be computed as follows, shown in the leftmost column below:

| | $2^V$ | $\mathbb{Z}_2^{\lvert V\rvert}$ | $E^{\lvert V\rvert}$ |
|---|---|---|---|
| $M_0$ | $\begin{pmatrix} \{N\} & & \\ & \{N,O\} & \\ & & \{N,O\} \end{pmatrix}$ | $\begin{pmatrix} \overset{L\,N\,O\,S}{\square\blacksquare\square\square} & & \\ & \square\blacksquare\blacksquare\square & \\ & & \square\blacksquare\blacksquare\square \end{pmatrix}$ | $\begin{pmatrix} E_{0,1} & & \\ & E_{1,2} & \\ & & E_{2,3} \end{pmatrix}$ |
| $M_1$ | $\begin{pmatrix} \{N\} & \varnothing & \\ & \{N,O\} & \{L\} \\ & & \{N,O\} \end{pmatrix}$ | $\begin{pmatrix} \square\blacksquare\square\square & \square\square\square\square & \\ & \square\blacksquare\blacksquare\square & \blacksquare\square\square\square \\ & & \square\blacksquare\blacksquare\square \end{pmatrix}$ | $\begin{pmatrix} E_{0,1} & E_{0,2} & \\ & E_{1,2} & E_{1,3} \\ & & E_{2,3} \end{pmatrix}$ |
| $M_2 = M_\infty$ | $\begin{pmatrix} \{N\} & \varnothing & \{S\} \\ & \{N,O\} & \{L\} \\ & & \{N,O\} \end{pmatrix}$ | $\begin{pmatrix} \square\blacksquare\square\square & \square\square\square\square & \square\square\square\blacksquare \\ & \square\blacksquare\blacksquare\square & \blacksquare\square\square\square \\ & & \square\blacksquare\blacksquare\square \end{pmatrix}$ | $\begin{pmatrix} E_{0,1} & E_{0,2} & E_{0,3} \\ & E_{1,2} & E_{1,3} \\ & & E_{2,3} \end{pmatrix}$ |

The same procedure can be translated, without loss of generality, into the bit domain ($\mathbb{Z}_2^{|V|}$) using a lexicographic nonterminal ordering, however $M_\infty$ in both $2^V$ and $\mathbb{Z}_2^{|V|}$ represents a decision procedure, i.e., $\big[S \in M_\infty[0,3]\big] \Leftrightarrow \big[M_\infty[0,3,3] = \blacksquare\big] \Leftrightarrow \big[A(\sigma) \neq \varnothing\big]$. Since $M_\infty[0,3] = \{S\}$, we know there is at least one $\sigma' \in A(\sigma)$, but neither $M_\infty$ in $2^V$ nor $\mathbb{Z}_2^V$ lets us recover a witness.

To witness $\sigma' \in A(\sigma)$, we can translate the matrix exponential to the provenance semiring. We first define $X \otimes Z = [X_2 \cdot Z_1, \varnothing, \varnothing, X_1 \cdot Z_0]$ and $X \oplus Z = [X_i \vee Z_i]_{i \in [0,|V|)}$, mirroring $\oplus, \otimes$ from the powerset domain. Since the preterminals $O, N$ can only occur on the superdiagonal, they may be safely ignored by $\otimes$. To solve for $M_\infty$, we proceed by first computing $E_{0,2}, E_{1,3}$:

$$\begin{aligned} E_{0,2} &= E_{0,j} \cdot E_{j,2} = E_{0,1} \otimes E_{1,2} \\ &= [L \in E_{0,2}, \varnothing, \varnothing, S \in E_{0,2}] \\ &= [O \in E_{0,1} \cdot N \in E_{1,2}, \varnothing, \varnothing, N \in E_{0,1} \cdot L \in E_{1,2}] \\ &= [E_{0,1,2} \cdot E_{1,2,1}, \varnothing, \varnothing, E_{0,1,1} \cdot E_{1,2,0}] \end{aligned}$$

$$\begin{aligned} E_{1,3} &= E_{1,j} \cdot E_{j,3} = E_{1,2} \otimes E_{2,3} \\ &= [L \in E_{1,3}, \varnothing, \varnothing, S \in E_{1,3}] \\ &= [O \in E_{1,2} \cdot N \in E_{2,3}, \varnothing, \varnothing, N \in E_{1,2} \cdot L \in E_{2,3}] \\ &= [E_{1,2,2} \cdot E_{2,3,1}, \varnothing, \varnothing, E_{1,2,1} \cdot E_{2,3,0}] \end{aligned}$$

Now we solve for the corner entry $E_{0,3}$ by dotting the first row and last column, which yields:

$$\begin{aligned} E_{0,3} &= E_{0,j} \cdot E_{j,3} = (E_{0,1} \otimes E_{1,3}) \oplus (E_{0,2} \otimes E_{2,3}) \\ &= [E_{0,1,2} \cdot E_{1,3,1} \vee E_{0,2,2} \cdot E_{2,3,1}, \varnothing, \varnothing, E_{0,1,1} \cdot E_{1,3,0} \vee E_{0,2,1} \cdot E_{2,3,0}] \end{aligned}$$

Since we only care about $E_{0,3,3} \Leftrightarrow [S \in E_{0,3}]$, we can ignore the first three entries and solve for:

$$\begin{aligned} E_{0,3,3} &= E_{0,1,1} \cdot E_{1,3,0} \vee E_{0,2,1} \cdot E_{2,3,0} \\ &= E_{0,1,1} \cdot (E_{1,2,2} \cdot E_{2,3,1}) \vee E_{0,2,1} \cdot \varnothing \\ &= E_{0,1,1} \cdot E_{1,2,2} \cdot E_{2,3,1} \big( = [N \in E_{0,1}] \cdot [O \in E_{1,2}] \cdot [N \in E_{2,3}] \big) \\ &= 1 \cdot \{+, \times\} \cdot \{0, 1\} \end{aligned}$$

Finally, to recover a witness, we can simply $\texttt{choose}\big(1 \cdot \{+, \times\} \cdot \{0, 1\}\big)$ (see Lemma 3.4).

### 4.3 Repair as intersection

We are now ready to consider the general case of syntax repair, in which case the edit locations are not localized but can occur anywhere inside the snippet. In this case, we construct a lattice of all possible edit paths up to a fixed distance. This structure is called a Levenshtein automaton.

As the original construction defined by Schulz and Mihov [56] contains cycles and $\varepsilon$-transitions, we propose a variant which is $\varepsilon$-free and acyclic. Furthermore, we adopt a symbolic form that supports infinite alphabets and simplifies the description to follow. Illustrated in Fig. 3 is an example of a small Levenshtein automaton recognizing $\mathcal{L}(L(\sigma : \Sigma^5, 3))$. Unlabeled arcs accept any terminal from the alphabet, $\Sigma$. Equivalently, this transition system can be viewed as a kind of proof system within an unlabeled lattice. The following construction is equivalent to Schulz and Mihov's original Levenshtein automaton, but is more amenable to our purposes as it does not contain any $\varepsilon$-arcs, and instead uses skip connections to recognize consecutive deletions of varying lengths.

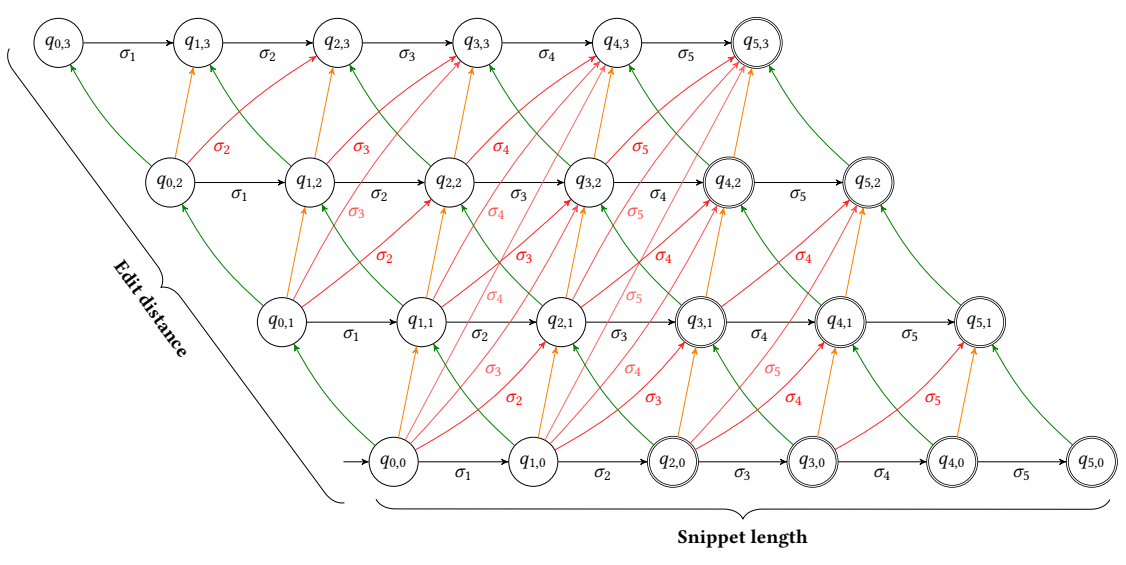

Fig. 3. Levenshtein NFA recognizing $\mathcal{L}(L(\sigma : \Sigma^5, 3))$.

$$\frac{s \in \Sigma \quad i \in [0, n] \quad j \in [1, d_{\max}]}{(q_{i,j-1} \overset{s}{\rightarrow} q_{i,j}) \in \delta} \qquad \frac{s \in \Sigma \quad i \in [1, n] \quad j \in [1, d_{\max}]}{(q_{i-1,j-1} \overset{s}{\rightarrow} q_{i,j}) \in \delta}$$

$$\frac{i \in [1, n] \quad j \in [0, d_{\max}]}{(q_{i-1,j} \overset{\sigma_i}{\rightarrow} q_{i,j}) \in \delta} \qquad \frac{d \in [1, d_{\max}] \quad i \in [d+1, n] \quad j \in [d, d_{\max}]}{(q_{i-d-1,j-d} \overset{\sigma_i}{\rightarrow} q_{i,j}) \in \delta}$$

$$\frac{}{q_{0,0} \in I}\ \textsc{Init} \qquad \frac{q_{i,j} \in Q \qquad |n - i + j| \leq d_{\max}}{q_{i,j} \in F}\ \textsc{Done}$$

Each type of arc plays a specific role. handles insertions, handles substitutions and handles deletions of one or more terminals. Let us consider some illustrative cases.

```
f . [ x )   f .   x )   f . ( x )   . + ( x )   f . ( x ;   [ , x y ]   [   , x y ]
f . ( x )   f . ( x )   f . (   )       ( x )   f *   x ;   [ x , y ]   [ x ,   y ]
```

Note that the same patch can have multiple Levenshtein alignments. Done constructs the final states, which are all states accepting strings $\sigma'$ whose Levenshtein distance $\Delta(\sigma, \sigma') \leq d_{\max}$.

To avoid creating a parallel bundle of arcs for each insertion and substitution point, we instead decorate each arc with a symbolic predicate, accepting or rejecting $\sigma_i$. To distinguish this symbolic variant from the original construction, we highlight the modified rules in orange below.

$$\frac{i \in [0, n] \quad j \in [1, d_{\max}]}{(q_{i,j-1} \overset{[\neq \sigma_{i+1}]}{\rightarrow} q_{i,j}) \in \delta} \qquad \frac{i \in [1, n] \quad j \in [1, d_{\max}]}{(q_{i-1,j-1} \overset{[\neq \sigma_i]}{\rightarrow} q_{i,j}) \in \delta}$$

$$\frac{i \in [1, n] \quad j \in [0, d_{\max}]}{(q_{i-1,j} \overset{[=\sigma_i]}{\rightarrow} q_{i,j}) \in \delta} \qquad \frac{d \in [1, d_{\max}] \quad i \in [d+1, n] \quad j \in [d, d_{\max}]}{(q_{i-d-1,j-d} \overset{[=\sigma_i]}{\rightarrow} q_{i,j}) \in \delta}$$

Using a symbolic predicate in the construction of the NFA avoids the unnecessary creation of $2(|\Sigma| - 1) \cdot |\sigma| \cdot d_{\max}$ arcs and reduces the representation size of the resulting automaton, but does not affect the underlying semantics. Since negation is only permitted over terminals and never propagates further, complementation can be struck from the usual RE axioms (Theorem 3.2).

As a concrete example, suppose we have the string $\sigma = \texttt{())}$ and wish to balance the parentheses. We will initially have the Levenshtein automaton, $A$, depicted in Fig. 4. To check for non-emptiness, we will perform the following procedure. Suppose we have a CNF CFG, $G' = \{S \to LR, S \to LF, S \to SS, F \to SR, L \to \texttt{(}, R \to \texttt{)}\}$ and let us assume an ordering $S < F < L < R$ of $V$.

First, we need to order the automaton states by increasing longest-path distance from $q_0$. One approach would be to topologically sort the adjacency matrix. While some form of sorting is unavoidable for arbitrary ANFAs, if we know ahead of time that our structure is a Levenshtein automaton, we can simply enumerate its state space by increasing Manhattan distance from the origin. So, a valid ordering on $Q$ would be $q_{00}, q_{01}, q_{10}, q_{11}, q_{20}, q_{21}, q_{30}, q_{31}$. Now, we want to compute whether $[\mathcal{L}(G') \cap \mathcal{L}(A) \neq \varnothing]$.

Under such an ordering, the adjacency matrix takes an upper triangular form and becomes the template for the initial parse chart, $M_0$ (Fig. 7). Each entry of this chart corresponds to a vector of expressions $E^{|V|}$ with at least one expression denoting a nonempty language. Likewise, the reachability matrix signifies a subset of state pairs which can participate in the language intersection. The adjacency and reachability matrices will always cover the expression vectors of the initial and final parse charts, respectively. In other words, we may safely ignore absent $\langle q, q' \rangle$ pairs in the reachability matrix, as these state pairs definitely cannot participate in the intersection.

From the reachability matrix we can construct the parse chart via matrix exponentiation. We note that n-step reachability constrains n-step parseability, i.e., $\sum_{i=0}^{n} A^i[q, q'] = \square \vdash M_n[q, q', v] = \square$, thus we can avoid substantial work via memoization. In this example, since $M_\infty[q_{00}, q_{31}, S] = \blacksquare$, this implies that $\mathcal{L}(A) \cap \mathcal{L}(G') \neq \varnothing$, hence $\text{LED}(\sigma, G) = 1$. Using the same matrix, we will then perform a second pass to construct regular expressions representing finite languages for each nonempty constituent. Once again, we can skip $\langle q, q', v \rangle$ entries when $M_\infty[q, q', v] = \square$ to hasten fixedpoint convergence.

Just as before, we will define $\oplus, \otimes$ using the provenance semiring defined in Thm. 3.2. Finally, we will repeat the matrix exponential, using $M_\infty$ in the binary domain as a guide, where $M_\infty[q, q', v] = \square \vdash \mathcal{L}(E[q, q', v]) = \varnothing$. This allows us to construct the regular expression tree for $S_\cap = q_{00}Sq_{20} \vee q_{00}Sq_{31}$ shown in Fig. 9. Once constructed, decoding becomes simply a matter of invoking $\texttt{choose}(S_\cap)$ (Lemma 3.4) with a suitable prior, which we will return to in § 6. In this case, there are only a few choices, but in general, there can be a vast multitude.

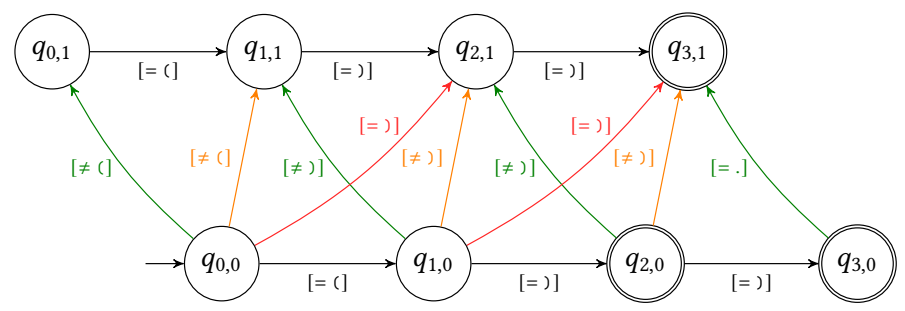

Fig. 4. Simple Levenshtein automaton.

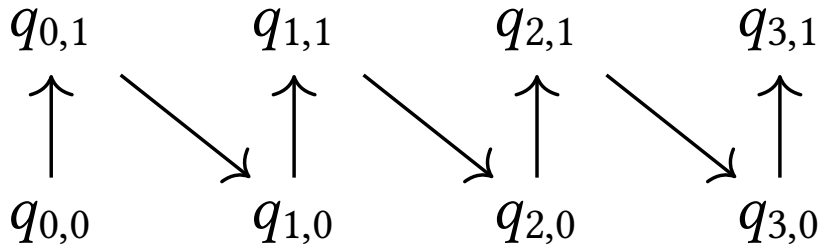

Fig. 5. Pairing function over $L(\sigma : \Sigma^3, 1)$.

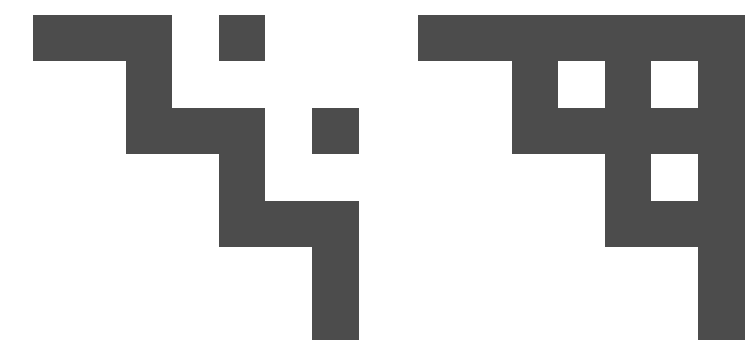

Fig. 6. Adjacency and reachability matrix.

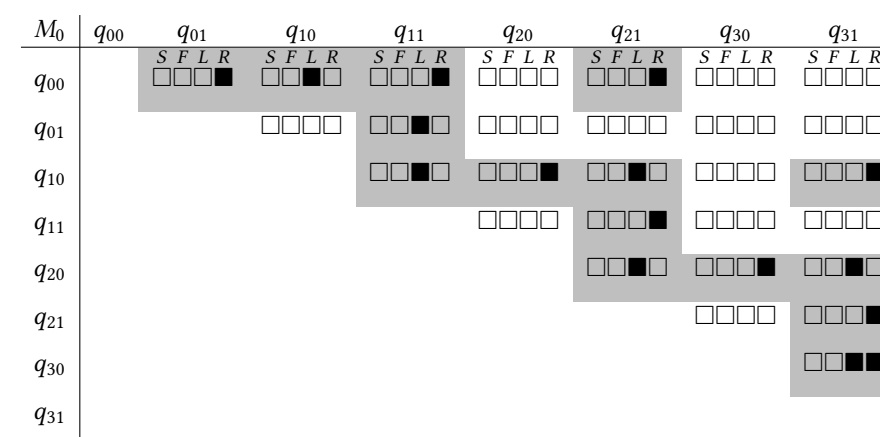

Fig. 7. Initial parse chart configuration.

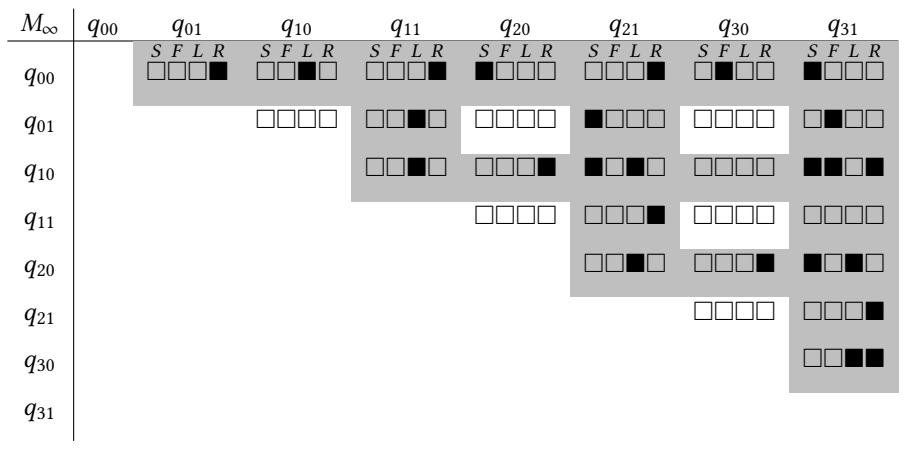

Fig. 8. Final parse chart configuration.

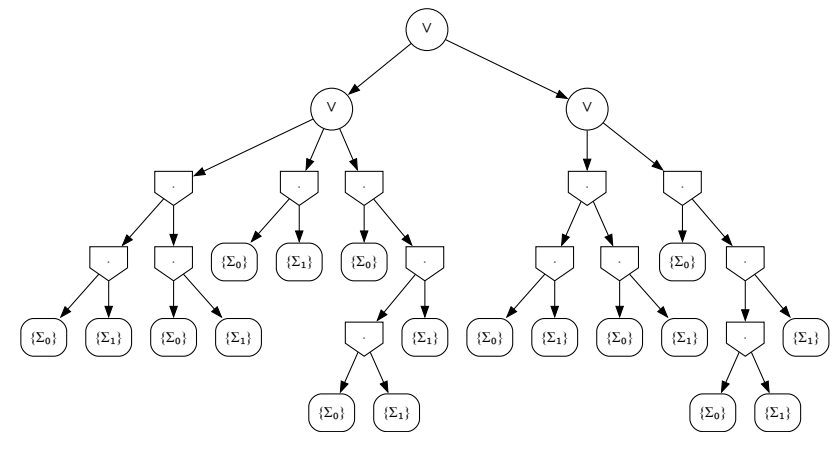

Fig. 9. Regular expression denoting $\mathcal{L}(G_\cap)$.

## 5 MEASURING THE LANGUAGE INTERSECTION

We will now attempt to put a probability distribution over the language intersection. We shall start with a few cursory but illustrative approaches, then proceed towards a more refined solution.

### 5.1 Mode collapse

Ordinarily, one might think to train a top-down PCFG sampler using a treebank of well-formed code snippets, however this method is highly degenerate in the finite case, exhibiting poor sample diversity. Consider an illustrative pathological case for top-down ancestral (TDA) sampling:

$$G = \left\{ S \rightarrow A\,B \left(\frac{10^5 - 1}{10^5}\right),\ S \rightarrow C\,C \left(\frac{1}{10^5}\right),\ A \rightarrow a\,(1),\ B \rightarrow b\,(1),\ C \rightarrow a \left(\frac{1}{26}\right) \mid \ldots \mid z \left(\frac{1}{26}\right) \right\}$$

Such a sampler will almost always yield $ab$, but most of $\mathcal{L}(G)$ is concealed in the hidden branch, $S \rightarrow CC$. Though a contrived example, it illustrates why TDA sampling is unviable: our sampler should match the true distribution over the finite CFL, not the PCFG's local approximation thereof.

### 5.2 Exact enumeration

To correct for mode collapse, a brute force solution would be to simply generate every tree. While the whole set can be materialized in some cases when the intersection language is small, this strategy is clearly suboptimal in the worst-case. Nevertheless, it is useful for checking completeness.

Theorem 5.1 (Enumeration). *We will first denote the number of unique trees in a regex as* $|e|$*:*

$$|e| : E \rightarrow \mathbb{N} = \begin{cases} 1 & \textit{if } e \in \Sigma \\ |x| \times |z| & \textit{if } e = x \cdot z \\ |x| + |z| & \textit{if } e = x \vee z \end{cases} \tag{4}$$

*To enumerate, we construct a bijection between trees and integers, then invoke* $\bigcup_{i=0}^{|e|-1} \left\{\texttt{enum}(e, i)\right\}$*:*

$$\texttt{enum}(e, n) : E \times \mathbb{N} \rightarrow \Sigma^* = \begin{cases} e & \textit{if } e \in \Sigma \\ \texttt{enum}\big(x, \lfloor \frac{n}{|z|} \rfloor\big) \cdot \texttt{enum}\big(z,\ n \bmod |z|\big) & \textit{if } e = x \cdot z \\ \texttt{enum}\big((x, z)_{\min(1, \lfloor \frac{n}{|x|} \rfloor)}, n - |x| \min(1, \lfloor \frac{n}{|x|} \rfloor)\big) & \textit{if } e = x \vee z \end{cases} \tag{5}$$

This can be converted to a uniform sampler by drawing integers without replacement using a pseudorandom number generator, however, if $|e|$ is very large, `enum` can fail to capture modes.

### 5.3 The problem with ambiguity

The main problem with Thm. 5.1 is that it counts distinct trees, which overcounts the number of words, $|\mathcal{L}(G_\cap)|$. Since the Levenshtein automaton can be ambiguous, uniform tree sampling causes certain repairs to be overrepresented, resulting in a pernicious bias. Consider, for example,

Lemma 5.2. *If the FSA,* $A_\varnothing$*, is ambiguous, then the intersection grammar,* $G_\cap$*, can be ambiguous.*

Proof. Let $\ell$ be the language defined by $G = \{S \rightarrow LR, L \rightarrow \texttt{(}, R \rightarrow \texttt{)}\}$, where $A_\varnothing = L(\sigma, 2)$, the broken string $\sigma$ is `)(`, and $\mathcal{L}(G_\cap) = \ell \cap \mathcal{L}(A_\varnothing)$. Then $\mathcal{L}(G_\cap)$ contains a word that can be parsed with two different trees: `)()` $\Rightarrow S_\cap \rightarrow q_{00}Lq_{21}\ q_{21}Rq_{22}$, and `()` $\Rightarrow S_\cap \rightarrow q_{00}Lq_{11}\ q_{11}Rq_{22}$. ☐

We would expect the underlying sample space to be a proper set, *not* a multiset. If we are to sample repairs uniformly at random, ambiguity cannot be permitted. To obtain the exact normalizing constant and disambiguate $\ell_\cap$, we require a different representation which will now be described.

### 5.4 Disambiguation

To count the number of distinct repairs, we will need to convert $G_\cap$ to an automaton. Since $\mathcal{L}(G_\cap)$ is finite, it must be regular a fortiori. Recalling the definition for an NFA, $\langle Q, \Sigma, \delta, q_\alpha : Q, F \subseteq Q\rangle$, and star-free regex, $e \rightarrow \Sigma \mid e \cdot e \mid e \vee e$, we will proceed by structural induction on the regex:

$$N(e) = \begin{cases} \Big\langle \{q_\alpha, q_\omega\} \;,\; \{q_\alpha \overset{e}{\rightarrow} q_\omega\} \;,\; q_\alpha \;,\; \{q_\omega\} \Big\rangle & \text{if } e \in \Sigma \\ \Big\langle Q_x \cup Q_z \;,\; \{q \overset{s}{\rightarrow} q_{\alpha z} \mid (q \overset{s}{\rightarrow} q_\omega^{\in F_x}) \in \delta_x\} \cup \delta_x \cup \delta_z \;,\; q_{\alpha x} \;,\; F_z \Big\rangle & \text{if } e = x \cdot z \\ \left\langle \begin{matrix} Q_x \cup \{q_{\alpha e}\} \cup \\ Q_z \cup \{q_{\omega e}\} \end{matrix} \;,\; \begin{matrix} \{q_{\alpha e} \overset{s}{\rightarrow} q \mid (q_{\alpha x, \alpha z} \overset{s}{\rightarrow} q) \in \delta_{x,z}\} \cup \delta_x \cup \\ \{q \overset{s}{\rightarrow} q_{\omega e} \mid (q \overset{s}{\rightarrow} q_\omega^{\in F_{x,z}}) \in \delta_{x,z}\} \cup \delta_z \end{matrix} \;,\; q_{\alpha e} \;,\; \{q_{\omega e}\} \right\rangle & \text{if } e = x \vee z \\ \text{- - - - - - - - - - - - or - - - - - - - - - - - -} & \\ \Big\langle Q_x \cup Q_z \cup \{q_{\alpha e}\} \;,\; \{q_{\alpha e} \overset{s}{\rightarrow} q \mid (q_{\alpha x, \alpha z} \overset{s}{\rightarrow} q) \in \delta_{x,z}\} \cup \delta_x \cup \delta_z \;,\; q_{\alpha e} \;,\; F_x \cup F_z \Big\rangle & \text{if } e = x \vee z \end{cases}$$

Though less conventional than Thompson's construction, $N(e)$ avoids the creation of unnecessary $\varepsilon$ arcs. And while slightly more verbose, we find the topology induced by the first version of the $\vee$ case to be more favorable for minimization. Continuing with our running example from § 4.3, we will use Brzozowski's algorithm [10] to construct the unique minimal DFA, $D_\cap^* \equiv_{\mathcal{L}} G_\cap$:

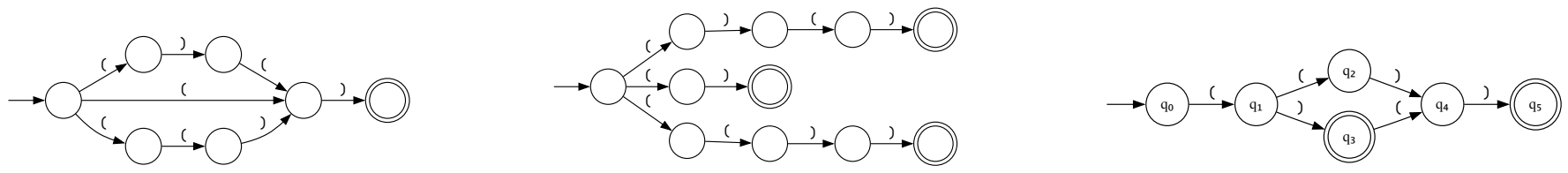

Fig. 10. FSA for $\mathcal{L}\big(L(\texttt{()}, 1)\big) \cap \mathcal{L}(G')$ (a) with or (b) without $\vee$-merging, and then (c) post-minimization.

Since $\mathcal{L}(G_\cap)$ is necessarily finite, we can infer that the corresponding DFA is acyclic and thus representable as an upper triangular adjacency matrix under a topological ordering of $\delta$. For any such DFA, we can ascertain the size of its language by counting walks from $q_\alpha$ to $q_\omega \in F$. Letting $A$ be the adjacency matrix for $D_\cap^*$, i.e., $A[q, q'] = |\{a \in \Sigma \mid \delta(q, a) = q'\}|$, the number of words it recognizes is given via the transfer matrix method [26], that is,

$$C(A, q_\alpha, F) : \mathbb{N}^{|Q| \times |Q|} \times Q \times 2^Q \rightarrow \mathbb{N} = \sum_{q_\omega \in F} (I - A)^{-1}[q_\alpha, q_\omega] = \sum_{q_\omega \in F} \sum_{i=0}^{|Q|-1} A^i[q_\alpha, q_\omega] \tag{6}$$

Plugging in powers of the adjacency matrix for the DFA shown in Fig. 10.(c), we arrive at the total:

$$(I - A)^{-1} = \quad I + A \quad + \quad A^2 \quad + \quad A^3 \quad + \quad A^4 \tag{7}$$

$$= \begin{pmatrix} 1 & 1 & & & & \\ & 1 & 1 & 1 & & \\ & & 1 & & 1 & \\ & & & 1 & 1 & \\ & & & & 1 & 1 \\ & & & & & 1 \end{pmatrix} + \begin{pmatrix} & & 1 & 1 & & \\ & & & & 2 & \\ & & & & & 1 \\ & & & & & 1 \\ & & & & & \\ & & & & & \end{pmatrix} + \begin{pmatrix} & & & & 2 & \\ & & & & & 2 \\ & & & & & \\ & & & & & \\ & & & & & \\ & & & & & \end{pmatrix} + \begin{pmatrix} & & & & & 2 \\ & & & & & \\ & & & & & \\ & & & & & \\ & & & & & \\ & & & & & \end{pmatrix} \tag{8}$$

$$= \begin{pmatrix} 1 & 1 & 1 & \underline{1} & 2 & \underline{2} \\ & 1 & 1 & 1 & 2 & 2 \\ & & 1 & & 1 & 1 \\ & & & 1 & 1 & 1 \\ & & & & 1 & 1 \\ & & & & & 1 \end{pmatrix} \text{ therefore, } \big|\mathcal{L}(D_\cap^*)\big| = C\big(A, q_0, \{q_3, q_5\}\big) = \underline{1} + \underline{2} = 3. \tag{9}$$

Note the model counting problem for arbitrary REs is strictly harder than deciding intersection nonemptiness as it requires determinization, however, weak bounds may be obtained by applying $C$ to the FSA generated by $N(e)$ or by direct analysis of $e$. While the inequality $C_{D_\cap^*} \leq C_{N(e)} \leq |e|$ will hold, the bounds provided by the latter approximations may be vacuous, whereas $C_{D_\cap^*}$ is exact. $D_\cap^*$ may be enumerated by standard methods [26] and sampled uniformly with a suitable PRNG.

## 6 DECODING AND RERANKING

The construction above gives us an exact finite language, $\ell_\cap = \mathcal{L}(G) \cap \mathcal{L}\big(L(\underset{\sim}{\sigma}, d)\big)$, whose contents are precisely the valid words within $d$ edits of $\underset{\sim}{\sigma}$. What remains is a ranking problem: among the words that are valid and nearby, which should be shown first? In selecting these repairs, our primary considerations will be (1) latency and (2) fidelity – within a few hundred milliseconds, we must produce a set that anticipates with high probability the author's intended repair. We separate this task into two subproblems: *support* and *ranking*. The support criterion must ensure no member of $\ell_\cap$ is lost and ranking ensures probability mass is maximized by the top-$k$ results.

### 6.1 Derivative-guided exact decoding

Let $e_\cap$ be the star-free regular expression produced by Theorem 3.2, so $\mathcal{L}(e_\cap) = \ell_\cap$. For a residual $e$, define $\mathsf{Done}(e) \iff \texttt{EOS} \in \mathcal{L}(e)$. A left-to-right decoder keeps a priority queue of triples $\langle \sigma, e, s\rangle$, where $\sigma$ is a prefix, $e = \partial_\sigma e_\cap$ is the residual and $s$ is a running total cost, which maintains order. Initially it is $[\langle \varepsilon, e_\cap, 0\rangle]$. Popping an item from the queue applies one of two possible cases,

$$\begin{aligned} \langle \sigma, e, s\rangle &\longmapsto \Big\langle \sigma a, \partial_a e, s + c(a \mid \underset{\sim}{\sigma}, \sigma)\Big\rangle && (a \in \texttt{next}(e)),\\ \langle \sigma, e, s\rangle &\longmapsto \Big\langle \mathsf{Done}(e), s + c(\texttt{EOS} \mid \underset{\sim}{\sigma}, \sigma)\Big\rangle && (\mathsf{Done}(e)), \end{aligned}$$

and a repair is emitted only when its completed item is popped. Here $c$ is a finite nonnegative action cost; when no explicit end-of-string model is used, take $c(\texttt{EOS} \mid \underset{\sim}{\sigma}, \sigma) = 0$.

Lemma 6.1 (Derivative decoder support). *Assume $\ell_\cap$ is finite, every legal action has finite cost, and the decoder never prunes a queue element. Then every emitted string lies in $\ell_\cap$, and every string in $\ell_\cap$ is eventually emitted exactly once.*

Proof. The invariant $e = \partial_\sigma e_\cap$ holds initially and is preserved by every ordinary transition. A completed item for $\sigma$ is created only when $\mathsf{Done}(e)$, i.e., when $\varepsilon \in \mathcal{L}(\partial_\sigma e_\cap)$, hence $\sigma \in \mathcal{L}(e_\cap) = \ell_\cap$. Conversely, if $\tau = a_1 \cdots a_n \in \ell_\cap$, then every strict prefix $a_1 \cdots a_i$ has nonempty residual language, so $a_{i+1} \in \texttt{next}(\partial_{a_1 \cdots a_i} e_\cap)$, and the final residual is nullable. The prefix tree of a finite language is finite, so all corresponding open and completed items are eventually popped. Each concrete prefix has a unique parent in this tree, so each word has exactly one completed item. ☐

If $c = -\log P$ for positive action probabilities, all costs are nonnegative and the queue is Dijkstra search over the finite prefix tree. Completed repairs are then emitted in nondecreasing total cost, or equivalently nonincreasing full-string probability. A cutoff preserves soundness but not full support, yielding an anytime top-$k$ prefix decoder. The remaining design choice is the scoring oracle, which controls how quickly probable repairs are reached.

### 6.2 Fast Markov retrieval

The simplest possible low-latency decoder is a smoothed $n$-gram model. For order $c$, let $h_c(\sigma)$ be the last $c - 1$ symbols of $\sigma$, padded on the left with $\texttt{BOS}$, and let $C(h, a)$ count occurrences of next token $a$ after history $h$ in valid programs. Writing $\Sigma_{\texttt{EOS}} = \Sigma \cup \{\texttt{EOS}\}$, we use

$$P_c(a \mid \sigma) = \frac{C\big(h_c(\sigma), a\big) + \lambda}{\sum_{b \in \Sigma_{\texttt{EOS}}} C\big(h_c(\sigma), b\big) + \lambda|\Sigma_{\texttt{EOS}}|}, \qquad a \in \Sigma_{\texttt{EOS}}.$$

Next-token decoding is effectively free: the derivative decoder can query this distribution with essentially zero latency. Its state update is simply $(\sigma, e, s) \longmapsto \big(\sigma a, \partial_a e, s - \log P_c(a \mid \sigma)\big)$, so it can often saturate small intersections and quickly sample very large intersections without replacement.

The primary weakness here is fidelity. The Markov state is the explicit suffix $h_c(\sigma)$, so $h_c(\sigma) = h_c(\tau) \implies P_c(\cdot \mid \sigma) = P_c(\cdot \mid \tau)$. This conflates stylometric properties that a good repair model should be able to distinguish, e.g., delimiter balance, indentation, and binding structure. While they provide a good latency baseline, $n$-gram approximations are generally too coarse and, by themselves, incapable of modeling the conditional probability $P(\sigma \mid \underset{\sim}{\sigma})$.

### 6.3 Constrained LLM decoding

A natural high-fidelity baseline keeps the exact admissibility layer from Lem. 6.1, but replaces the Markov table with a conditional repair model. We train a small GPT-style completer on paired broken and repaired snippets. For a training pair $\langle \underset{\sim}{\sigma}, \sigma^\star \rangle$, form

$$x = (\mathtt{BOS}, \underset{\sim}{\sigma}_1, \ldots, \underset{\sim}{\sigma}_m, \mathtt{MOS}, \sigma^\star_1, \ldots, \sigma^\star_n, \mathtt{EOS}).$$

Only the repair suffix contributes to the left-to-right loss. If $b = m + 2$ is the position of `MOS`, then

$$\mathcal{L}_{\text{GPT}}(\theta) = -\sum_{\langle \underset{\sim}{\sigma}, \sigma^\star \rangle} \sum_{t=b}^{|x|-1} \log P_\theta(x_{t+1} \mid x_{\leq t}).$$

The broken snippet is therefore visible as context, while the induced distribution is

$$P_\theta(\sigma \mid \underset{\sim}{\sigma}) = \left( \prod_{i=1}^{|\sigma|} P_\theta(\sigma_i \mid \underset{\sim}{\sigma}, \sigma_{<i}) \right) P_\theta(\mathtt{EOS} \mid \underset{\sim}{\sigma}, \sigma).$$

At decoding time, let $e = \partial_\sigma e_\cap$. The admissible actions are

$$\mathsf{Adm}(e) = \mathtt{next}(e) \cup \begin{cases} \{\mathtt{EOS}\} & \text{if } \mathsf{Done}(e), \\ \varnothing & \text{otherwise.} \end{cases}$$

We use $\mathsf{Adm}(e)$ as a hard support mask. For an ordinary terminal $a \in \mathtt{next}(e)$, the decoder pushes

$$\left\langle \sigma a, \partial_a e, s - \log P_\theta(a \mid \underset{\sim}{\sigma}, \sigma) \right\rangle;$$

when $\mathsf{Done}(e)$, it pushes $\left\langle \mathtt{done}(\sigma), s - \log P_\theta(\mathtt{EOS} \mid \underset{\sim}{\sigma}, \sigma) \right\rangle$ onto the queue. If $a = \mathtt{EOS}$, the prefix $\sigma$ is emitted as a complete repair. Therefore, the LLM only orders prefixes whose residual language is nonempty. It cannot introduce strings outside $\ell_\cap$, and if the queue is run to saturation, the support remains exactly the same finite language as in Lemma 6.1.

The advantage of this decoder is fidelity: it conditions jointly on the broken snippet, the entire repair prefix, and long-range syntactic or stylistic cues that are invisible to a fixed suffix model. The disadvantage is throughput. Each expansion requires a transformer forward pass before the grammar mask can discard inadmissible successors. A width-$B$ beam can reduce cost, but it loses support completeness unless $B \geq |\ell_\cap|$, which is exhaustive search in disguise. In practice, constrained LLM decoding is therefore a useful high-fidelity baseline for scoring $P(\sigma \mid \underset{\sim}{\sigma})$ over $\ell_\cap$, but a poor first-stage retriever under real-time latency constraints, and as we will show in § 7.4, underperforms a dedicated reranker even with a generous test-time latency advantage.

Instead, we will consider a sound regular overapproximation which provides a much faster next-token scoring interface for retrieval without compromising completeness with respect to $\ell_\cap$.

### 6.4 Weighted Nederhof overapproximation

Our goal is to construct a finite-state decoder whose next-token queries are as cheap as an $n$-gram lookup, but whose state can summarize more than a fixed suffix. To achieve this, we turn to Nederhof's regular approximation. Importantly, this approximation is not used to generate repairs directly. Repairs are still generated from the exact expression $e_\cap$. The approximation supplies only a lightweight language prior used to order the next-token distribution.

Nederhof [45] constructs $\alpha_h$ with $\mathcal{L}(G) \subseteq \mathcal{L}(\alpha_h)$ by truncating the call history of a recursive transition network (RTN). Let

$$\mathcal{R} = \{R_A = (Q_A, \delta_A, \iota_A, F_A) \mid A \in V\}, \qquad \delta_A \subseteq Q_A \times (\Sigma \cup V \cup \{\varepsilon\}) \times Q_A$$

be the RTN for the CFG $G = \langle \Sigma, V, P, S \rangle$. An arc $p \xrightarrow{B} q$ is a call to the component $R_B$. For an unfolding depth $h$, the automaton $\alpha_h$ is obtained by expanding calls for the first $h$ call levels:

$$p \xrightarrow{B} q \quad \rightsquigarrow \quad p \xrightarrow{\varepsilon} \iota'_B \quad \text{and} \quad f'_B \xrightarrow{\varepsilon} q \quad (f'_B \in F'_B).$$

At depth $h$, remaining calls are connected through shared copies of their callee automata rather than unfolded again. Finally, we remove $\varepsilon$-arcs, determinize, and minimize, obtaining a DFA $\alpha_h$. The construction forgets enough call-stack information to become regular, and therefore may accept strings not in $\mathcal{L}(G)$ but does not exclude any string generated by $G$.

As an example, let us consider the grammar $S \to \mathtt{x} \mid S\,S \mid \{\,S\,\}$, which has the following RTN:

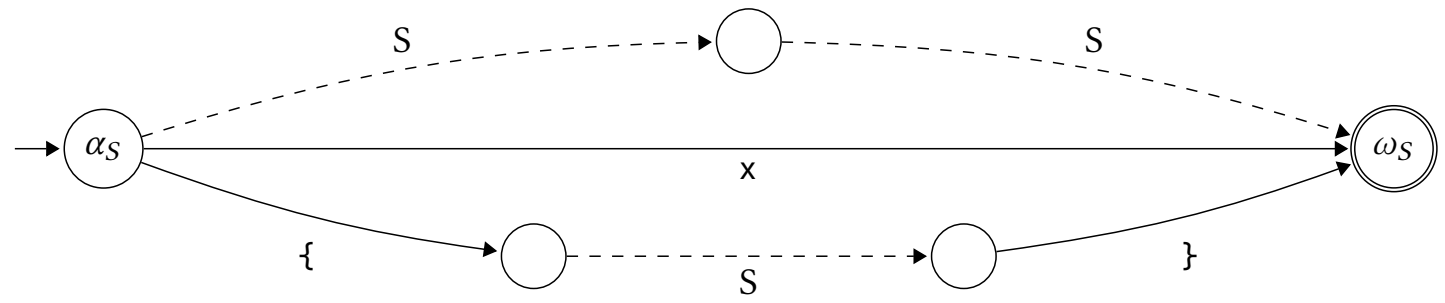

Then we replace $p \xrightarrow{S} q$ arcs with a copy of the RTN and recurse up to depth h. On this grammar, the effect of the depth-bounded unfolding can be visualized by the $\varepsilon$-removed, determinized, and minimized DFAs for $1 \leq h \leq 4$:

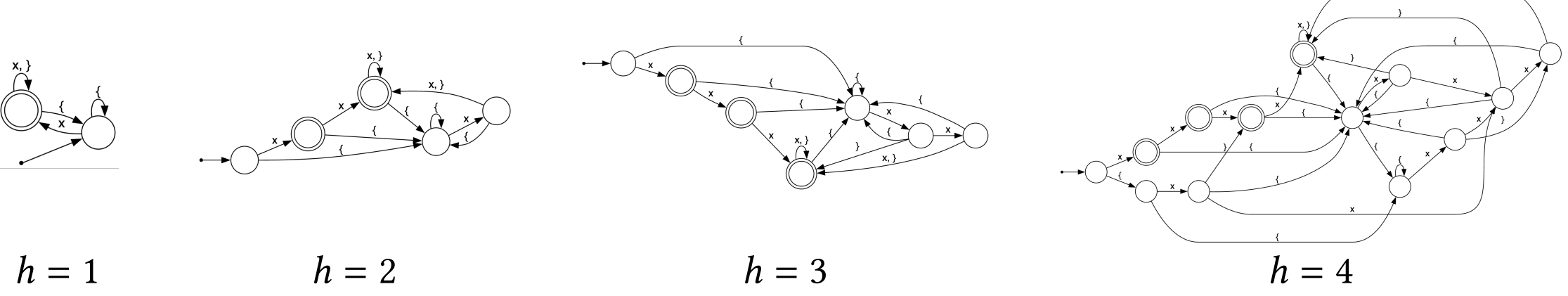

Fig. 11. Minimal DFA generated by the Nederhof construction for $h \in [1, 4]$.

We will now show that the Nederhof overapproximation does not compromise completeness:

Lemma 6.2 (Nederhof support cover). *For every unfolding depth* $h \geq 0$, $\mathcal{L}(G) \subseteq \mathcal{L}(\alpha_h)$. *Consequently, for every* $\underset{\sim}{\sigma}$ *and edit radius* $d$, $\ell_\cap = \mathcal{L}(G) \cap \mathcal{L}\big(L(\underset{\sim}{\sigma}, d)\big) \subseteq \widehat{\ell}_{h,\cap} = \mathcal{L}(\alpha_h) \cap \mathcal{L}\big(L(\underset{\sim}{\sigma}, d)\big)$.

Proof. Every derivation in $G$ has an equivalent well-nested accepting RTN run. Terminal and $\varepsilon$-arcs are copied directly. At or beyond depth $h$, a call enters the appropriate shared callee copy and may choose the return edge leading to its actual continuation, so every stack-disciplined run remains an accepting path. Sharing also permits mismatched returns, which accounts for the overapproximation. Since $\varepsilon$-removal and determinization preserve language, $\mathcal{L}(G) \subseteq \mathcal{L}(\alpha_h)$; intersecting both sides with $\mathcal{L}\big(L(\underset{\sim}{\sigma}, d)\big)$ preserves inclusion. □

Once the Nederhof DFA is constructed, we can attach weights without changing its support. Let $\alpha_{h,\hat{\theta}}$ be the weighted version of $\alpha_h$, with deterministic transition function $\delta$, initial state $q_0$, positive transition weights $w(q, a) > 0$ on all arcs, and positive final weights $w_F(q) > 0$ at accepting states. For an accepting run $q_i = \delta(q_{i-1}, \sigma_i)$, define

$$P_{\hat{\theta}}(\sigma_1 \cdots \sigma_n) = \left( \prod_{i=1}^{n} w(q_{i-1}, \sigma_i) \right) w_F(q_n),$$

and set $P_{\hat{\theta}}(\sigma) = 0$ when $\sigma \notin \mathcal{L}(\alpha_h)$. Positivity and Lemma 6.2 give the weighted support cover

$$\ell_\cap \subseteq \widehat{\ell}_{h,\cap} \subseteq \operatorname{supp}(P_{\hat{\theta}}).$$

During exact decoding, a queue element carries both the residual expression $e = \partial_\sigma e_\cap$ and the WDFA state $q_\sigma$. An extension $a \in \texttt{next}(e)$ advances the scorer by the query update $q_{\sigma a} = \delta(q_\sigma, a)$ and adds the cost $-\log w(q_\sigma, a)$, a constant-time lookup operation. If $\mathsf{Done}(e)$, termination adds $-\log w_F(q_\sigma)$. Transitions are only pushed when admitted by $\texttt{next}(e)$, *not* the WDFA. Thus strings in $\widehat{\ell}_{h,\cap} \setminus \ell_\cap$ may influence the learned state representation, but are never emitted as repairs. This gives the intended first-stage tradeoff: latency comparable to an $n$-gram model, with a richer regular abstraction and a formal guarantee that every valid repair has nonzero probability.

By default, we choose $h = 2$ and learn weights over the resulting DFA. In our Python repair experiments, this produces a DFA with 2,115 states and 82,257 transitions, to which we fit locally-normalized weights on a corpus of Python snippets; further details can be found in Appendix E.

### 6.5 LaTeR reranking

The WDFA prior is fast, but it is still an unconditional language model: $P_{\hat{\theta}}(\sigma)$, not $P(\sigma \mid \underset{\sim}{\sigma})$, which is what we need. The final challenge will be to restore this missing conditional information by reranking a small exact candidate list. Let $C_K^\infty(\underset{\sim}{\sigma}) = \mathrm{TopK}_{P_{\hat{\theta}}}(\ell_\cap)$ denote the exact top-$K$ members of $\ell_\cap$ under the WDFA prior. In the implementation, the WDFA-guided decoder is stopped after a budget $T$ or after exhausting $\ell_\cap$, whichever occurs first, yielding an anytime retrieved list

$$C_{K,T}(\underset{\sim}{\sigma}) \subseteq \ell_\cap, \qquad |C_{K,T}(\underset{\sim}{\sigma})| \leq K.$$

The Levenshtein-aligned Transformer Reranker (LaTeR) scores only this retrieved set. Let $\mathrm{LA}(\underset{\sim}{\sigma}, \sigma)$ be a minimum-cost Levenshtein alignment, encoded as token-level edit features indicating which parts of the candidate match, substitute, insert, or correspond to deletions in the broken snippet. We encode it as a sequence over $(\Sigma \cup \{\bot\})^+ \times (\Sigma \cup \{\bot\})^+ \times \{\texttt{0}, \texttt{S}, \texttt{I}, \texttt{D}\}^+$. The reranker uses a query encoder and an alignment-aware candidate encoder,

$$E_\phi(\underset{\sim}{\sigma}) : \Sigma^n \to \mathbb{R}^p, \qquad E'_\phi\big(\sigma, \mathrm{LA}(\underset{\sim}{\sigma}, \sigma)\big) : (\Sigma \times \mathbb{Z}_4)^{[m,n]} \to \mathbb{R}^p,$$

and uses an MLP : $\mathbb{R}^p \times \mathbb{R}^p$ to assign a scalar relevance score to each candidate repair,

$$r_\phi(\underset{\sim}{\sigma}, \sigma) = \mathrm{MLP}_\phi\Big(E_\phi(\underset{\sim}{\sigma}), E'_\phi\big(\sigma, \mathrm{LA}(\underset{\sim}{\sigma}, \sigma)\big)\Big).$$

For a training pair $\langle \underset{\sim}{\sigma}, \sigma^\star \rangle$ with $\sigma^\star \in \ell_\cap$, define $D_K(\underset{\sim}{\sigma}) = C_{K,T}(\underset{\sim}{\sigma}) \cup \{\sigma^\star\}$. If retrieval misses it, the true repair is inserted into the candidate list for subsequent reranking. This forced insertion occurs only during training and never during inference. We train with the tempered listwise objective,

$$\mathcal{L}_{\mathrm{LaTeR}}(\phi) = -\sum_{\langle \underset{\sim}{\sigma}, \sigma^\star \rangle} \log \frac{\exp\big(r_\phi(\underset{\sim}{\sigma}, \sigma^\star)/\tau\big)}{\sum_{\sigma \in D_K(\underset{\sim}{\sigma})} \exp\big(r_\phi(\underset{\sim}{\sigma}, \sigma)/\tau\big)}.$$

At inference time, LaTeR reorders the retrieved list and returns the reranked top-k repairs:

$$\ell_\cap \xrightarrow{\text{WDFA retrieval}} C_{512,10^7}(\underset{\sim}{\sigma}) \xrightarrow{\text{LaTeR}} \mathrm{Top}_k^{r_\phi}\, C_{512,10^7}(\underset{\sim}{\sigma}).$$

This factorization separates support from fidelity. The CFL–regular intersection determines the exact candidate language $\ell_\cap$ and the Nederhof WDFA is a sound support overapproximation for retrieving promising candidates quickly, which LaTeR then reorders using the broken snippet and edit alignment. While reranking can improve presentation order, it cannot repair retrieval failure: if the true repair is not in $C_{K,T}(\underset{\sim}{\sigma})$, LaTeR will not be able to recover it.

This pipeline, consisting of language intersection (§ 3), WDFA-guided retrieval (§ 6.4) and LaTeR reranking (§ 6.5), forms the basis for the syntax repair tool used in the following experiments.

## 7 EVALUATION

We call our end-to-end syntax repair tool Tidyparse and consider the following research questions:

- **RQ 1**: What statistical properties do human repairs exhibit? (e.g., length, edit distance)
- **RQ 2**: How effective is Tidyparse at fixing syntax errors? (i.e., vs. Seq2Parse, BIFI, GPT-5.4)
- **RQ 3**: Which design choices are most significant? (e.g., decoding, reranking, parallelism)

We address **RQ 1** in § 7.2 by analyzing the distribution of natural code snippet lengths and edit distances, **RQ 2** in § 7.3 by comparing against three external syntax repair baselines, and **RQ 3** in § 7.4 by ablating various design choices and evaluating the impact on precision and latency.

### 7.1 Experimental setup

In the following set of experiments, we use syntax errors and fixes from the Python language. Python code snippets are abstracted as a sequence of lexical tokens using the official Python 3.8.11 parser, erasing alphanumeric identifiers and literals but retaining all other keywords. Accuracy is evaluated across a test set of pairwise errors and repairs by checking for lexical equivalence with the ground-truth repair, following the same methodology as Sakkas et al. (2022) [54].

We use the Precision@k statistic, which measures the frequency of the true repair appearing in the top-$k$ results, across a dataset of repair instances. Specifically, given a repair model, $R : \Sigma^* \longrightarrow (\Sigma^*)^t$ and a test set, $\mathcal{D}_{\text{test}}$, containing pairwise aligned errors ($\underset{\sim}{\sigma}$) and fixes ($\sigma^\star$), we define Precision@k as:

$$\text{Precision@k}(R) = |\mathcal{D}_{\text{test}}|^{-1} \sum_{\langle \underset{\sim}{\sigma}, \sigma^\star \rangle \in \mathcal{D}_{\text{test}}} \mathbb{1}\left[\sigma^\star \in R(\underset{\sim}{\sigma})_{1\ldots k}\right] \tag{10}$$

Our full dataset [64] consists of $2 \times 10^4$ naturally-occurring pairs of Python errors and human fixes from Stack Overflow, which we use to evaluate the precision of each model at blind recovery of the ground truth repair. From the Stack Overflow dataset, we filter for syntax errors shorter than 80 tokens and fewer than four lexical edits apart from the corresponding repair, then divide the remaining repairs into two disjoint sets: a training set ($\mathcal{D}_{\text{train}}$) of 4,586 repair instances and a test set ($\mathcal{D}_{\text{test}}$) of 2,238 repair instances, balanced across each length interval and edit distance, i.e., $\mathcal{D}_{\text{test}} = \{\langle \underset{\sim}{\sigma}, \sigma^\star \rangle \mid \lfloor |\underset{\sim}{\sigma}|/10 \rfloor \in [0, 8], \Delta(\underset{\sim}{\sigma}, \sigma^\star) \in [1, 3]\}$, each test bin containing at least 50 instances.

To train the reranker, we augment each instance in the training and test set with a list of repair confounders. Each instance consists of a tuple, $\langle \underset{\sim}{\sigma} : \bar{\ell}, \sigma^\star : \ell_\cap, \boldsymbol{\sigma} : \ell_\cap^{\leq 512} \rangle$, with a single syntax error ($\underset{\sim}{\sigma}$), the ground truth repair, ($\sigma^\star$), and up to 512 confounders ($\boldsymbol{\sigma}$) sampled without replacement from $\ell_\cap$ using our pretrained WDFA. We then train the reranker on $\mathcal{D}_{\text{train}}$ for 13,000 steps which takes $\sim$ 4 hours, and evaluate on $\mathcal{D}_{\text{test}}$. Hyperparameters are provided in Appendix E.

For our final repair procedure, we use the CNF Python grammar, $G_{\text{Python}}$ and let $d_{\max}$ be the smallest $d$ such that $\ell_\cap = \mathcal{L}(G) \cap \mathcal{L}\big(L(\underset{\sim}{\sigma}, d-1)\big)$ is nonempty. We decode $\ell_\cap$ with the same WDFA used during reranker training, and pass the top-512 results by WDFA probability to the transformer encoder, then finally rerank the top-512 by softmax probability and measure the Precision@k across repairs of differing length and edit distance in the test set.

We compare our method with three external baselines on the same test set: Seq2Parse, Break-It-Fix-It (BIFI) [66], and GPT-5.4 mini prompted for one tokenized repair. The Seq2Parse and BIFI experiments ran on a single Nvidia V100 GPU with 32 GB RAM. For Seq2Parse, we use the default pretrained model from commit `7ae0681` [3]. For BIFI, we use the Round 2 breaker and fixer from commit `ee2a68c`[4], the highest-performing model reported by the authors, with a variable-width beam search to control the number of predictions, and let the BIFI fixer model predict the top-$\{1, 2 \times 10^4\}$ repairs. Tidyparse runs on an Apple MacBook M4 Max with 128 GB RAM.

[3] https://github.com/gsakkas/seq2parse/tree/7ae0681f1139cb873868727f035c1b7a369c3eb9

[4] https://github.com/michiyasunaga/BIFI/tree/ee2a68cff8dbe88d2a2b2b5feabc7311d5f8338b

### 7.2 Dataset statistics

In the following experiments, we use a dataset of Python snippets consisting of 20,500 pairwise-aligned human errors and fixes from Stack Overflow [64]. We preprocess the dataset to lexicalize all code snippets, then filter by length and distance shorter than 80 lexical tokens and under four edits, i.e., with Levenshtein distance under four lexical edits $\left(|\Sigma| = 88, |\sigma| < 80, \Delta(\sigma, \sigma^\star) < 4\right)$. We depict the length, edit distance, normalized edit locations and stability profile in Fig. 12.

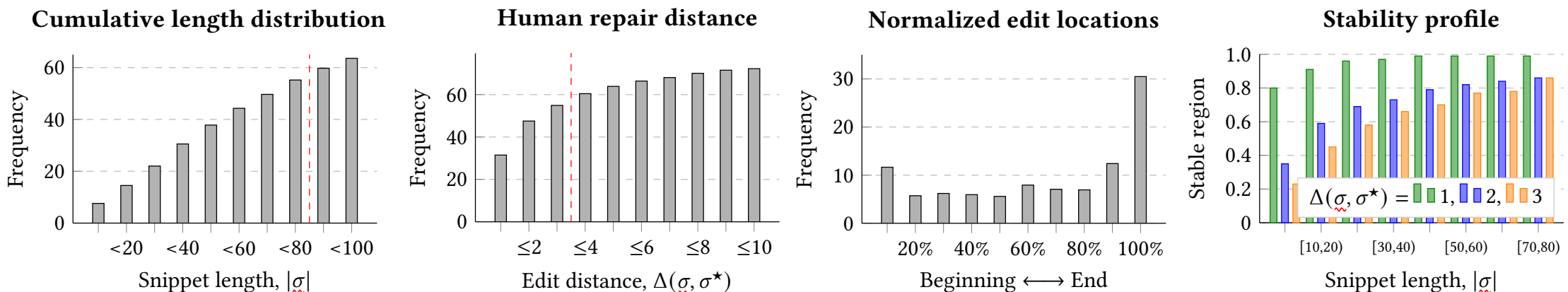

Fig. 12. Repair statistics across the Stack Overflow dataset. Tidyparse can handle about 90% in under ∼1s and ∼4 GB. Larger repairs and edit distances are possible, albeit requiring additional time and memory.

We observe that slightly over 6,800 code snippet pairs in the Stack Overflow dataset contain fewer than 80 tokens and four lexical edits, which are computationally feasible to process in a few hundred milliseconds. We also note a slight primacy or recency bias in the edit locations, evidenced by a large fraction of human repairs which modify the boundaries of the broken code snippet.

For the stability profile, we enumerate repairs for each syntax error and estimate the average fraction of all edit locations that were never altered by any repair in the $L\left(\sigma, \Delta(\sigma, \sigma^\star)\right)$-ball. For example, on average roughly half of the string is stable for 3-edit syntax repairs in the $[10 - 20)$ token range, whereas 1-edit repairs of the same length could modify only $\sim$ 10% of all locations. For a fixed edit distance, we observe an overall decrease in the relative freedom of edit location with increasing length, which intuitively makes sense, as the repairs are more heavily constrained by the surrounding context and their locations grow more concentrated relative to the entire string.

For an intuition about the size of the language intersections involved in syntax repair, volumetric analysis will be helpful, particularly in understanding the influence of snippet length and edit distance on language intersection volume. To measure the intersection volume we will form the $L\left(\sigma, \Delta(\sigma, \sigma^\star)\right)$ automaton, intersect it with the Python grammar, then automatize the resulting regular expression and finally compute the DFA transfer matrix using the method described in § 5.4 to obtain the exact volume. For a given (error, fix) pair, this tells us how many repairs of equal or lesser distance exist in the Python language. Plotting intersection volume across the full dataset (Fig. 13), we observe a strong positive correlation with the Levenshtein margin and a mild correlation with snippet length. Fully materializing $\ell_\cap$ is typically only feasible if we extend the Levenshtein radius up to one edit beyond the language edit distance (LED) $\left(\text{i.e., } d_{\max} \leq \text{LED}(\sigma)\right.$[5]$\left.+1\right)$ after which it grows too large to exhaustively generate and must be sampled. Across all snippets where $\Delta(\sigma, \sigma^\star) < 4$, approximately 54% matched LED($\sigma$), 35% had an edit distance of LED($\sigma$) + 1 and 11% had a distance of LED($\sigma$) + 2.

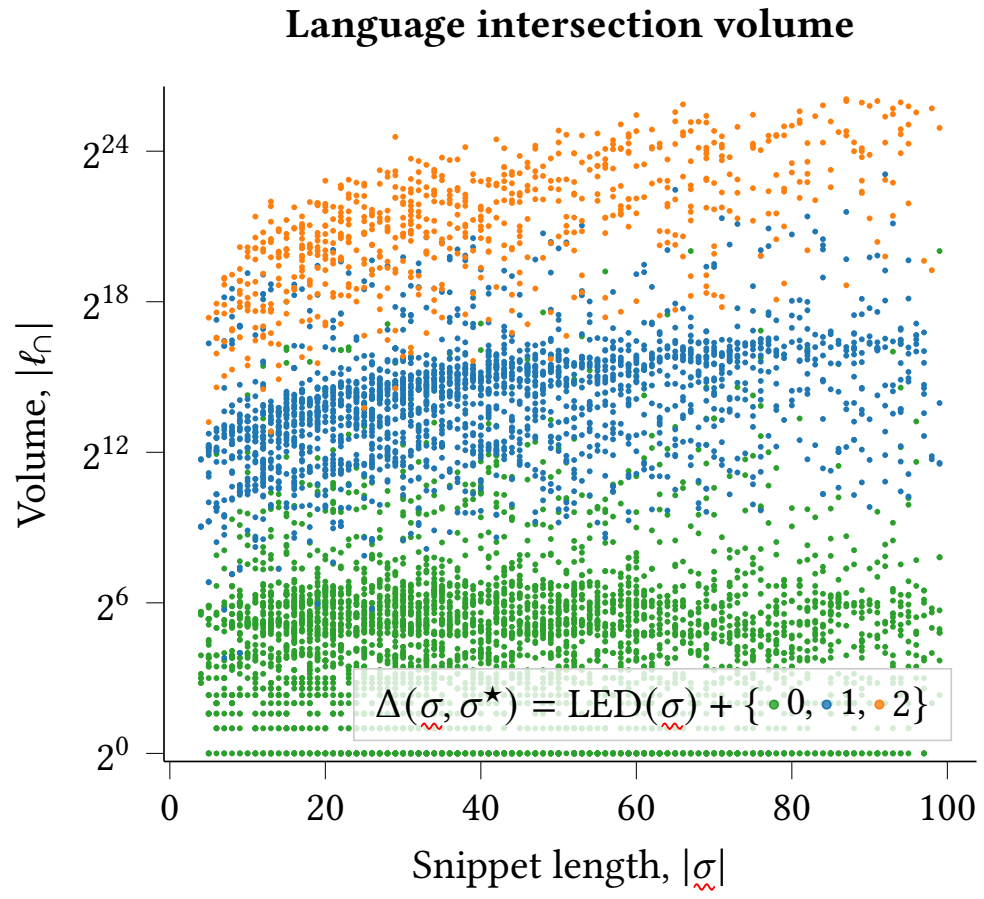

Fig. 13. Language volume versus snippet length and edit distance for Python repairs.

[5]Where LED($\sigma$) is shorthand for $\text{LED}(\sigma, \ell) = \min\left\{d_{\max} : \mathbb{N} \mid \mathcal{L}\left(L(\sigma, d_{\max})\right) \cap \ell \neq \varnothing\right\}$ with $\ell$ being the Python language.

### 7.3 Stack Overflow evaluation

For our first experiment, we measure the top-1 precision of our repair procedure at various lengths and Levenshtein distances, comparing Tidyparse against Seq2Parse and BIFI on the same test set. We also include a GPT-5.4 token-repair baseline using a short prompt followed by the same tokenized repair format. Each bin in the test set contains at least 50 distinct repairs sampled uniformly at random from the Stack Overflow dataset, none of which were present in the training set.

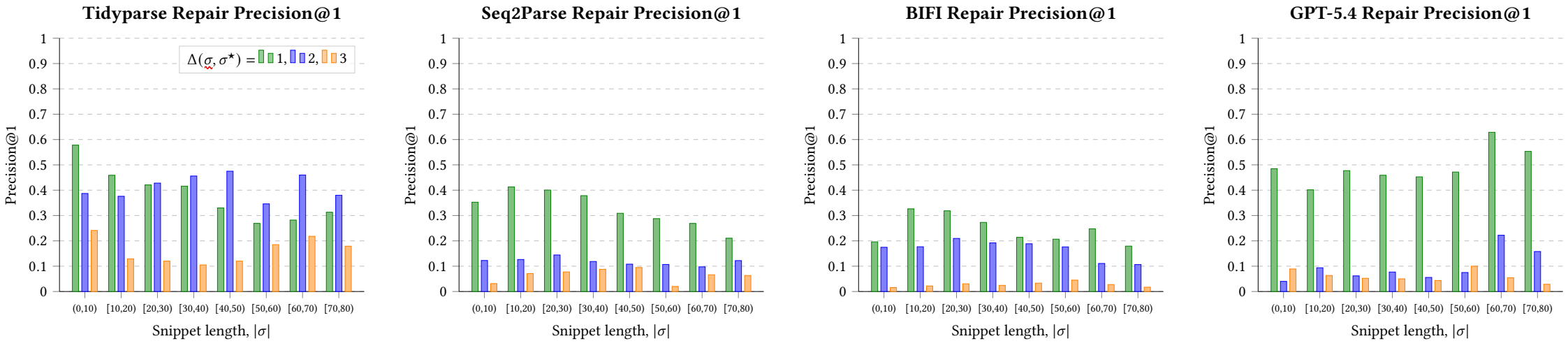

Fig. 14. Probability of the first recommendation matching the true repair for Tidyparse, Seq2Parse, BIFI, and GPT-5.4 across various lengths and Levenshtein distances.

As seen in Fig. 14, Tidyparse attains state-of-the-art top-1 repair precision by a significant margin among syntax-repair baselines. Precision does not monotonically decrease with edit distance, as Tidyparse's double-edit Precision@1 slightly outperforms single-edit Precision@1 across the test set. Although Seq2Parse outperforms BIFI by a lower margin, results are slightly mixed across edit distance. Although the GPT-5.4 probe is competitive on longer single-edit repairs, Precision@1 degrades sharply as edit distance increases. Similar to Fig. 13, the nonlinear correlation between edit distance and repair precision holds across all four models.

For the next experiment, we evaluate the BIFI model, giving it a generous compute and latency advantage with an unlimited time budget to sample $2 \times 10^4$ repairs, and compare the Precision@10 of our approach with a 10s timeout. As Tidyparse uses a WDFA for decoding, it can sample a much larger candidate set in the time allotted, but must use a transformer-based reranker after decoding to sort the top-512 repairs. Since the Seq2Parse reference implementation does not support sampling more than one repair, we do not compare its Precision@k for higher k values. The raw data from these experiments can be found in Appendix D.

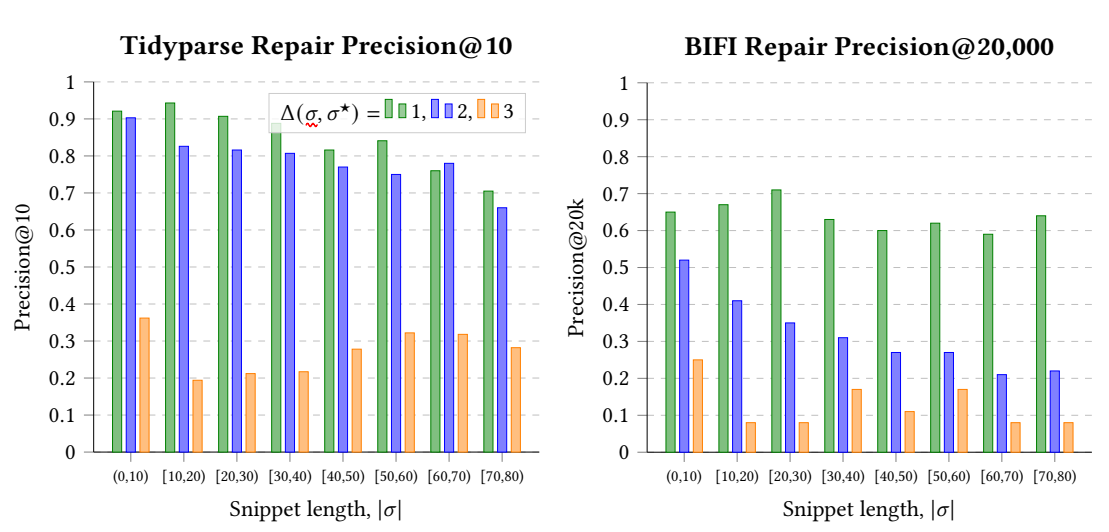

Fig. 15. Probability of the true repair being in the first ten Tidyparse repairs, and the first $2 \times 10^4$ BIFI repairs.

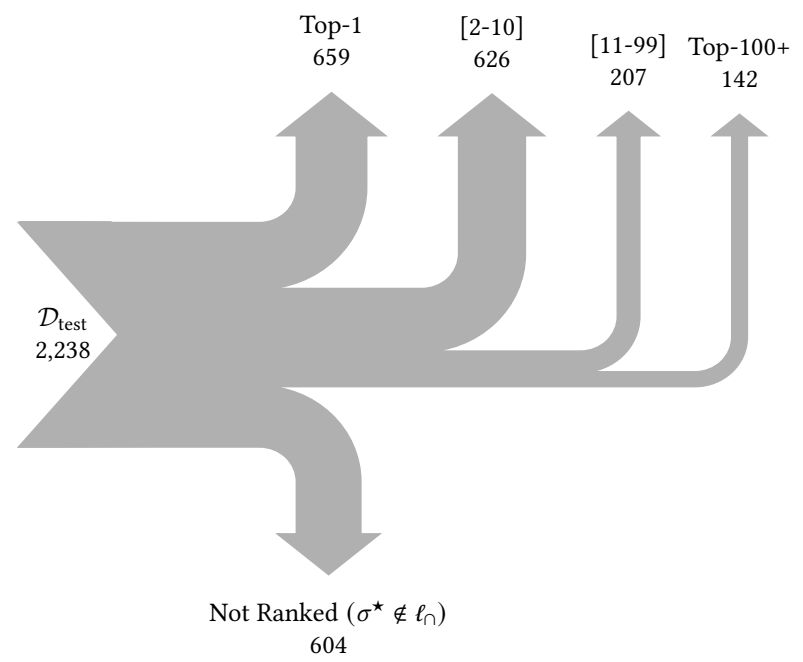

Fig. 16. Outcomes in the repair pipeline.

We present a Sankey diagram of the Tidyparse repair pipeline in Fig. 16. Across 2,238 test set repairs filtered by length and distance ($\lfloor|\sigma|/10\rfloor \in [0, 8], \Delta(\sigma, \sigma^\star) < 4$), we evaluated Tidyparse with a timeout of 10s and tracked individual repair outcomes. In 604 cases, the true repair was not contained in the language intersection and thus never sampled, in 1,634 cases the true repair was sampled, of which 659 cases the first prediction matched the true repair, in 1,285 cases, the true repair was in the top-10 results, and in the remaining 349 cases the true repair was drawn, but ranked lower than $10^{\text{th}}$ in the final results.

### 7.4 Internal evaluation

This ablation isolates the effect of scoring and reranking after the intersection expression has already been constructed. All methods search the same finite support, $\ell_\cap$, and validity is enforced before any statistical model is consulted. If derivative decoding is allowed to exhaust the prefix tree, every scorer eventually exposes the same complete repair set. Under a rank cutoff or timeout, however, the order in which $\ell_\cap$ is explored matters: a better first-stage scorer places the intended repair earlier in the queue, while a better reranker improves the final order of the retrieved list.

We compare LaTeR against three first-stage scoring models. The **4-gram** model is a Laplace-smoothed Markov baseline over valid snippets, chosen for its very low decoding latency. The **GPT-2** model is a left-to-right conditional repair model trained on paired broken and repaired snippets; it has access to the broken input, but is expensive to query inside the decoding loop. The **WDFA** model is a learned weighted finite-state prior whose support is a Nederhof overapproximation of the target grammar. It does not generate repairs by itself: the derivative decoder still draws candidates from the exact language $\ell_\cap$, while the WDFA supplies a constant-time score for ordering prefix expansions. Finally, **LaTeR**, a trained ranking model, reranks the top-512 WDFA-retrieved repairs produced by the WDFA.

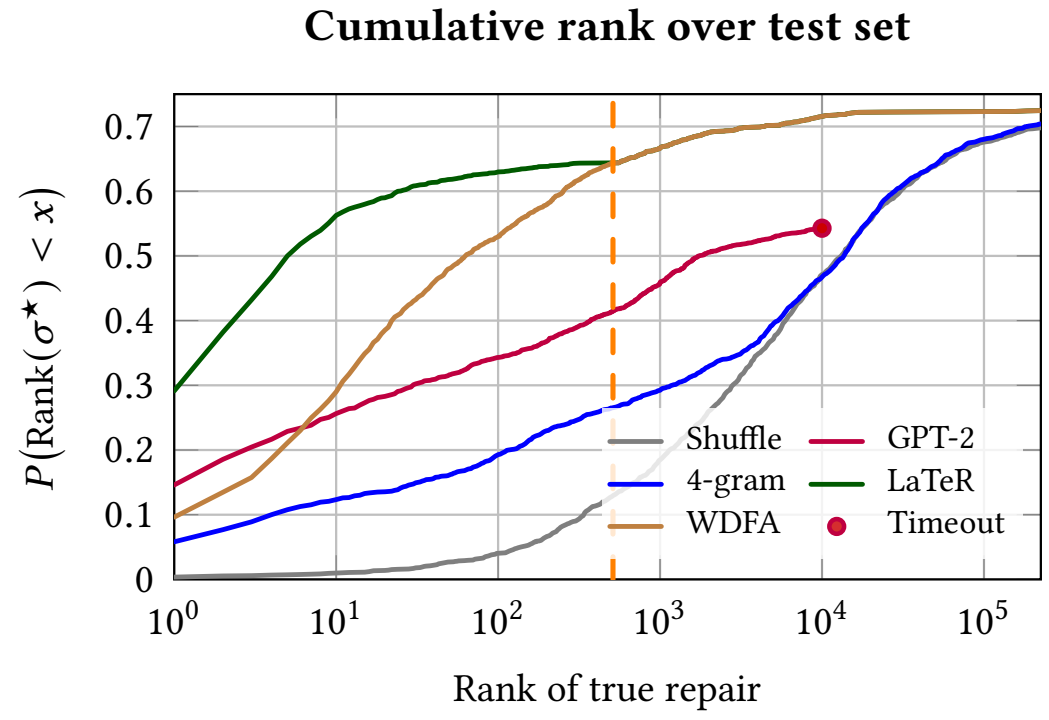

Fig. 17. Cumulative rank of the ground-truth repair under various scoring and reranking models.

Fig. 17 plots, for each method, the cumulative distribution $P(\text{Rank}(\sigma^\star) < x)$ of the ground-truth repair's rank across the test set. The higher the curve, the earlier the intended repair is found, on average, across the test set. The shuffle curve gives a random-ordering baseline, the dashed vertical line marks the 512-candidate handoff to LaTeR. The constrained GPT-2 decoder is terminated after sampling $\sim 10^4$ repairs, which can take several minutes per repair. The gap between the 4-gram and WDFA curves demonstrates the value of replacing a local suffix model with a learned regular-state scorer and the gain from WDFA retrieval to LaTeR reranking illustrates the additional value of conditioning on the broken snippet and the edit alignment with an expressive neural model. Since LaTeR only reorders candidates it has already received, it cannot alter syntactic validity or edit distance, nor can it recover a true repair that was not retrieved before the cutoff.

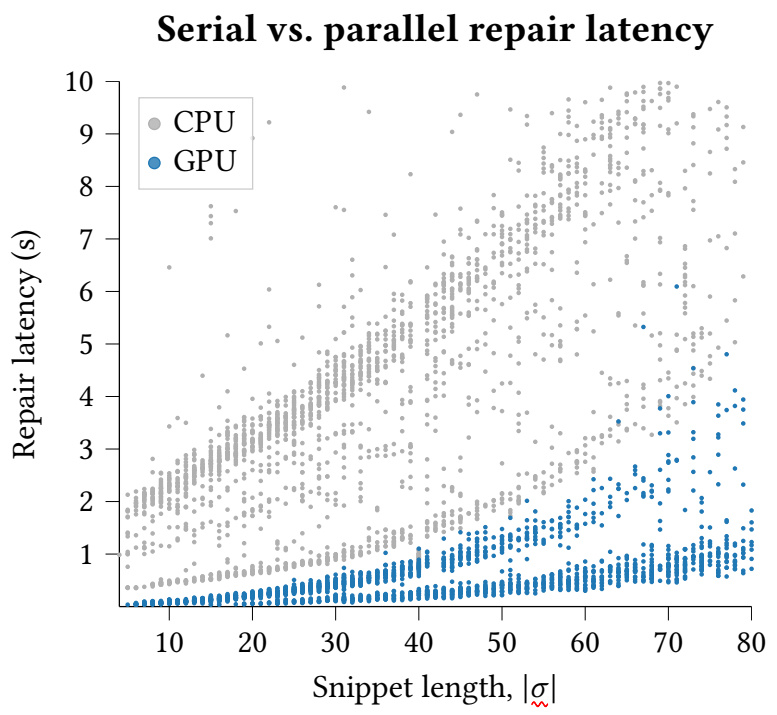

Fig. 18. End-to-end repair timings.

Finally, we investigate the impact of increased parallelism on repair throughput. To simulate a realistic editing scenario, we measure the end-to-end wall-clock runtime required to construct the Levenshtein automaton, form the intersection regex ($S_\cap$), decode and rerank the intersection language ($\ell_\cap$). Then, using our GPU implementation of the same algorithm (described in § A), we compare the individual repair timings across the same test set. Both the GPU and CPU versions run on the Apple M4 processor. As shown in Fig. 18, GPU execution has slightly lower variance and confers a 3-5× speedup across common repair scenarios. We conjecture the visible clustering effect is an artifact of thermal throttling on the M4, as repair instances in the same cluster share no salient properties in common such as LED or intersection size that would explain the observed timing variance. The GPU implementation finishes processing nearly all repairs within 10s and about 90% of all instances in under 1s; the CPU implementation may miss that target on longer or multi-edit repairs.

## 8 DISCUSSION

The main lesson we draw from § 7 is that it is feasible to significantly improve the precision of real-time syntax repair by incorporating syntactic constraints such as edit distance, then sampling and evaluating a large set of candidate repairs using a fast primary decoder and a more intensive secondary reranker. Though sample-efficient, direct transformer decoding comes at considerable cost to throughput, resulting in fewer repairs being discovered in a fixed amount of time.

Our approach uses a grammar and a high-throughput $n$-gram decoder to fetch the initial candidate repairs, then employs the transformer-encoder to rerank only the top-scoring repairs from the retrieved set. This allows us to repair errors in real-world programming languages and provides far more flexibility and controllability during the repair process, resulting in significantly higher precision on downstream repairs and ultimately, a smoother user experience.

Our primary insight leading to state-of-the-art precision is that repairs are typically concentrated near the center of a small Levenshtein ball, and by enumerating or sampling it carefully, then reranking repairs by naturalness, one can achieve significantly higher precision than single-shot neural repair. This is especially true for small-radii Levenshtein balls, where the intersection language is small enough to be completely enumerated and ranked. For larger radii, we can still achieve state-of-the-art precision by sampling a representative subset within a fixed timeout.

There is a clear tradeoff between latency and precision for any repair model. While existing neural syntax repair models scale poorly with additional time, Tidyparse is highly effective at exchanging more time for higher precision. We find that the Precision@10 of our method is competitive with BIFI's Precision@$(2 \times 10^4)$, while requiring only a fraction of the inference time. Unlike neural syntax repair models, Tidyparse can sample directly from the language specification, removing the possibility of hallucination. The emphasis on completeness is especially useful for discovering small or contextually improbable repairs, which are easily overlooked by neural models.

Although latency and precision are ultimately the deciding usability factors, repair throughput is a crucial intermediate factor to consider when evaluating the performance of a repair system. Even with a perfectly accurate reranker, if the correct repair is never retrieved, it will be for naught. By maximizing the total number of unique valid repairs, we increase the probability of retrieving natural repairs to give the reranker the best chance of surfacing them to the user.

One might be tempted to model syntax repair as a rejection sampling problem, but as Fig. 13 portrays, this strategy would be mistaken. Even if checking a single repair for validity takes just 1 ms, complete enumeration could take 24+ hours, and we have mere seconds at most. While rejection sampling has lower latency to find admissible repairs, it wastes a tremendous amount of computation and scales poorly with edit distance. It is far better to spend more computation upfront by performing the intersection in a way that avoids rejection and returns natural repairs.

Likewise, methods that rely on decoding large language models appear to face a similar dilemma. As shown in Fig. 15, even if we sample thousands of repairs from BIFI, an LLM specifically trained on syntax repair, it is possible to miss natural valid repairs of a given distance that would be easily found by an extensive $c$-gram search of $\ell_\cap$, plus reranking. This suggests completeness may be equally, if not more important, than sample efficiency for the purposes of evaluating candidate repairs. The rank CDF in Fig. 17 reinforces this point: lightweight scoring can surface many ground-truth repairs, and stronger conditional scorers improve their presentation order.

Taken together, these results provide strong evidence to support the central claim made in the introduction (§ 1): *existing syntax repair methods simply generate far too few repairs to be effective.* By extensively generating and evaluating a large quantity of repairs within a fixed edit distance, we show it is possible to predict the author's intent far more reliably, with greater precision and lower latency than competing methods which rely solely on transformer-based neural networks.

## 8.1 Limitations and future work

We identify three broad categories of limitations in Tidyparse and suggest directions for future work: naturalness, complexity, and toolchain integration.

*8.1.1 Naturalness.* First, Tidyparse does not currently support intersections between weighted CFGs and weighted finite automata, à la Pasti et al. [49]. This feature would allow us to put transition probabilities on the Levenshtein automaton corresponding to edit probability, then construct a weighted intersection grammar. With this, one could preemptively discard unlikely productions from $G_\cap$ to reduce the complexity of intersection in exchange for relaxed completeness. We also hope to explore alternate sampling strategies such as sequential Monte Carlo [38] and denoising diffusion models [6] with an inductive bias for structured sampling via Levenshtein edits.

The reranker is currently evaluated over lexical tokens, but we expect that a more precise ranking function could be constructed by incorporating names and numbers from the original source code and then scoring plain text. Furthermore, the decoder only considers each candidate repair $P_\theta(\sigma^\star)$ in isolation, returning the most probable candidates independent of the original error. This could be improved by conditioning the decoding on the broken sequence ($\underset{\sim}{\sigma}$), parser error message ($m$), original source ($s$), and possibly other contextual priors to assist sample efficiency.

*8.1.2 Complexity.* Latency can vary depending on several factors including string length, grammar size, and critically the Levenshtein edit distance. This can be an advantage because, without any contextual or statistical information, syntax and minimal Levenshtein edits are often sufficiently constrained to identify a small number of valid repairs. It is also a limitation because the admissible set expands rapidly with edit distance and the Levenshtein metric diminishes in usefulness without a very precise metric to discriminate natural solutions in the cosmos of equidistant repairs.

Space complexity increases sharply with edit distance and to a lesser extent with length. This can be partly alleviated with various encoding tricks and a more efficient GPU implementation, but the memory overhead is still considerable. Memory pressure can be attributed to engineering factors such as the grammar encoding, but is also an inherent challenge of language intersection. Therefore, managing the size of the intersection grammar by preprocessing the syntax and automaton and using an efficient representation are critical factors in scaling up our technique.

*8.1.3 Toolchain integration.* Program slicing is an important preprocessing consideration that has so far gone unmentioned. The current implementation expects pre-sliced code fragments, however in a more practical scenario, it would be necessary to leverage editor information to identify the boundaries of the repairable fragment. One solution would be to just use the line surrounding the caret position, however a more complete solution requires careful editor integration.

Lastly and perhaps most significantly, Tidyparse does not incorporate semantic constraints, so its repairs, whilst syntactically admissible, are not guaranteed to be type-safe, and must be filtered by some form of compiler or incremental type checker before being presented to the user. It may be possible to add a type-based semantic refinement to our language intersection, however this would require a more expressive grammatical formalism than CFGs naturally provide.

Extending language intersections to handle type error repair requires leaving the domain of syntax and entering the much more daunting world of semantics – here one must contend with difficult questions in mathematical logic and finite model theory. One direction would be to collapse these problems down to context-free languages using a finite model embedding à la Considine [18]. Another direction would be to increase the expressivity of the grammar, using something like conjunctive grammars [47]. A third approach would be to adopt the framework of contextual modal type theory, then study the behavior of Levenshtein edit distance on modal accessibility in weak substructural type systems like the Lambek calculus [50]. We leave this for future work.

## 9 RELATED WORK

Our work draws a threefold correspondence between well-known techniques in (1) formal language theory, (2) program analysis and (3) incremental decoding. We will first survey these topics, then turn our attention to machine learning, with which we compare and partly use for reranking.

### 9.1 Formal language theory

Context-free language (CFL) parsing is the well-studied problem of how to turn a string into a unique tree, with many different algorithms and implementations (e.g., shift-reduce, recursive-descent, LR). Many of those algorithms expect grammars to be expressed in a certain form (e.g., left- or right-recursive) or are optimized for a narrow class of grammars (e.g., regular, linear).

General CFL parsing allows ambiguity (non-unique trees) and can be formulated as a dynamic programming problem, as shown by Cocke-Younger-Kasami (CYK) [53], Earley [23] and others. These parsers have roughly cubic complexity with respect to the length of the input string.

As shown by Valiant [60], Lee [37] and others, general CFL recognition is in some sense equivalent to binary matrix multiplication, another well-studied combinatorial problem, known to be at worst subcubic. This reduction opens the door to a range of complexity-theoretic speedups to CFL recognition; however large constants tend to limit their practical applicability. In the parallel setting, Ruzzo [52] presents a polylogarithmic parallel parser for length-$n$ strings, however the overhead can be substantial. Brent and Goldschlager [9] independently show an approach requiring $\mathcal{O}(n^6)$ processors, which Mikhelson and Okhotin [43] later extend to weighted semirings. Instead, we opt for a naïve parallel CKY algorithm, which is linear time decidable on $\mathcal{O}(n^3)$ processors.

Bar-Hillel [7] proves the closure of CFLs under intersection with regular languages, but does not elaborate on how to construct the corresponding grammar. Salomaa [55] and Pasti et al. [49] provide helpful insights into constructing the intersection grammar, and Nederhof and Satta [46] specifically consider finite CFL intersections, but seem unaware of the connection to CFL reachability. Our work specializes Bar-Hillel intersections to Levenshtein automata in particular, and more generally acyclic automata using a refinement of Salomaa's construction [55] based on CFL reachability.

### 9.2 CFL reachability

Our contribution is closely related to the literature on CFL reachability. In brief, the CFL reachability problem seeks to determine, given an edge-labeled graph and distinguished vertex pair, $\langle v, v' \rangle$, whether there is a path, $v \rightsquigarrow v'$, whose concatenated edge labels are contained in the CFL. For a deeper overview, see Zhang and Su [67]. This problem has been known [51] for some time [35] to have broad applications to program analysis and as our work finds, to syntactic program repair.

From a complexity-theoretic perspective, the CFL reachability problem is known to be at worst subcubic [16] with polylogarithmic time factors. Koutris and Deep [36] present a fine-grained complexity analysis, with concurrent work by Istomina et al. [34] expanding on fine-grained reductions. Muravev and Grigorev explore how to accelerate this technique on a GPU [44].

Surprisingly absent from the literature on CFL reachability is a discussion of the Bar-Hillel construction, regular expressions for witnessability, or the use of Brzozowski's derivative for incremental decoding. Nor does the literature specifically consider finite intersection nonemptiness between CFLs and acyclic automata such as the Levenshtein NFA (§ 4.3). In Theorem 3.1 we give a constructive proof via a parallel matrix closure with a linear bound on the number of squaring rounds, borrowing the matrix multiplication technique from CFL reachability to build a star-free regular expression that we decode using the Brzozowski derivative. This technique sheds new light on the Bar-Hillel construction, and translates to a simple and efficient implementation which is fully compatible with autoregressive decoding techniques used in probabilistic language modeling.

### 9.3 Language equations

Language equations are a powerful tool for reasoning about formal languages and their inhabitants. First proposed by Ginsburg et al. [27] for the ALGOL language, language equations are essentially systems of inequalities with variables representing *holes*, i.e., unknown values, in the language or grammar. Solutions to these equations can be obtained using various fixpoint techniques, yielding members of the language. This insight reveals the true algebraic nature of CFLs and their cousins.

Being an algebraic formalism, language equations naturally give rise to a kind of calculus, vaguely reminiscent of Leibniz's and Newton's. First studied by Brzozowski [11, 12] and Antimirov [5], one can take the derivative of a language equation, which can be interpreted as a kind of continuation or language quotient, revealing the suffixes that complete a given prefix. This technique leads to an elegant family of algorithms for incremental parsing [1, 42] and regular expression matching [57, 61].

Recently Pasti et al. [48] introduce an elegant grammar-to-grammar transformation and next-token decoding technique that gives the exact next-token probabilities for a given PCFG, across all prefixes. This result is a surprising one, because the weighted algorithm was only recently discovered, despite PCFGs and autoregressive decoding techniques being used for over 70 years.

More concretely, we restrict our attention to language equations over CFLs whose variables coincide with edit locations in the source code of a computer program, and solutions correspond to syntax repairs. While prior work has studied the use of language equations for parsing [42], to our knowledge, they were never specifically considered for code completion or syntax error correction.

### 9.4 Syntax repair

In finite languages, syntax repair corresponds to spelling correction, a more restrictive and largely solved problem. Schulz and Mihov [56] construct a finite automaton that returns the nearest dictionary entry by Levenshtein edit distance. Though considerably simpler than syntax correction, their work shares similar challenges and offers insights for handling more general repair scenarios.

When a string is syntactically invalid, parsing grows more challenging. Like spelling, the problem is to find the minimum number of edits required to transform an arbitrary string into a syntactically valid one, where validity is defined as containment in a (typically) context-free language. Early work, including Irons [33] and Aho [2] proposes a dynamic programming algorithm to compute the minimum number of edits required to fix an invalid string. Prior work on error-correcting parsers only considers the nearest edit(s), and does not study edits of varying distance in the Levenshtein ball. Furthermore, the problem of repair is not generally well-posed, as there can be many valid solutions. We instead focus on maximum probability Levenshtein-CFL reachability, which seeks to find the most natural syntactically valid repair within a fixed Levenshtein distance.

Diekmann and Tratt [20] present a rule-based syntax repair tool that retrieves the complete set of minimum-cost repairs, but only works for deterministic CFLs, a proper subset of the CFL family which admit a linear-time parser. Their cost model is based on insertion and deletion, and does not consider probability or non-minimal edit distance. Tidyparse can handle arbitrary CFLs and generate repairs within an arbitrary edit distance, using a Levenshtein cost model.

Zhang et al. [68] introduce OrdinalFix, which uses CFL reachability to repair compiler errors, however their method only returns admissible repairs and not necessarily probable ones. As they do not consider the problem of maximum-probability repairs, nor use any form of ranking to sort the results by naturalness or probability, we do not compare with their work.

### 9.5 Decoding

Decoding is an important subproblem in machine translation, speech recognition, and sequence-to-sequence learning. Given a compressed encoding of some learned distribution, the goal is to find the maximum probability samples. A classic example is Viterbi decoding, used to find the most likely path through a hidden Markov model (HMM), a kind of weighted automaton.

In particular, we care about the problem of *top-$k$ decoding*, which attempts to find the exact or approximate $k$-most likely samples in order of decreasing probability. This is closely related to the $k$-best enumeration [25] problem, a carefully studied problem in graph theory and combinatorial optimization. An exact solution to this problem for large acyclic CFGs is often intractable, but we can approximate it using a beam search over subwords, then rerank top-scoring results.

A popular solution to $k$-best decoding in the NLP literature is a technique called cube-pruning [13, 32], which samples maximum probability paths through a hypergraph. We take inspiration from this technique and adapt it to the setting of constrained decoding from finite CFGs.

An alternate line of work originates from combinatorics [31] and Boltzmann sampling [22], which constructs a generating function for the language and samples it uniformly. This technique has applications to constraint satisfaction and model counting problems in formal languages.

A third approach would be to use some form of constrained decoding [40, 59, 63] such as sequential Monte Carlo to steer an autoregressive LLM, as proposed by Lew et al. [38, 39]. These techniques show promise for program repair, however, the question of whether to use left-to-right decoding or some other strategy is still unresolved in the language modeling community. For example, there is an emerging class of flow-based or structured denoising diffusion models [6] which starts from a noise distribution and iteratively denoises it by sampling one or more edits at random locations with each decoding step. Typical work focuses on audiovisual data, but very recent work by Havasi et al. [30] adapt this to the Levenshtein edit model for generating source code. Although these models do not yet use CFGs or consider language intersections, they are inherently more fault-tolerant than decoders which require expensive backtracking-style search.

### 9.6 Learning-based program repair methods

The last decade has seen a surge of progress in programming with large language models. That work is primarily based on methods from differential calculus and continuous optimization, leading to the so-called *naturalness hypothesis* [3], which suggests programming languages are not so different from natural ones. In contrast, PL theory takes the view that languages are essentially discrete sets governed by logical calculi. Programming, thus viewed, is more like a mathematical exercise in constraint satisfaction. These two approaches have more in common than would seem.

From an applied perspective, a number of gradient-based methods have been introduced to repair programming errors [4, 15, 21]. These approaches typically employ large language models (LLMs) and treat the problem as a sequence-to-sequence transformation. While capable of generating natural repairs, these models are susceptible to misgeneralization, costly to train, and challenging to customize thereafter. Furthermore, the generated repairs are not necessarily sound without additional filtering, and we observe the released models often hallucinate false-positive repairs.

Two prior works specifically address syntax repair, Break-It-Fix-It (BIFI) [66] and Seq2Parse [54]. BIFI adapts techniques from semi-supervised learning to generate synthetic errors in clean code and fixes them. This reduces the need for pairwise training data, but generalizes poorly to lengthy or out-of-distribution repairs. Seq2Parse combines a transformer-based model with an augmented version of the Earley parser to suggest error rules, but only suggests a single repair.

Recent work by Merrill et al. [41] and Chiang et al. [14] suggests that the issue with generalization may be more foundational: transformer-based language models, a popular class of neural language models used in program synthesis and repair, are fundamentally less expressive than context-free grammars, which formally describe the syntax of many programming languages. This suggests such models, despite their useful approximation properties, are ill-suited for the task of end-to-end syntax repair. Yet, as our work demonstrates, they can be useful for resolving ambiguity between valid repairs of differing probability or reranking a set of repair candidates drawn from a CFL.

Using the RASP model from Weiss et al. [62], Yang et al. [65] characterize the expressive power of hard-attention transformers in terms of star-free regular expressions. While their work uses formal language theory to investigate language learnability and structural priors in transformers, it shares fruitful connections to CFL reachability, which, similar to RASP, treats matrix multiplication as a kind of programmable interface into which various state tracking problems and static analysis tasks can be compiled and analyzed in a common linear algebraic framework.

## 10 CONCLUSION

Our work, while a case study on syntax repair, is part of a broader line of inquiry in program synthesis that investigates how to weave formal language theory and machine learning into helpful programming tools for everyday developers. In some ways, syntax repair serves as a test bench for integrating learning and language theory, as it lacks the intricacies of type-checking and semantic analysis, but is still rich enough to be an interesting challenge. By starting with syntax repair, we hope to lay the foundation for more organic hybrid approaches to program synthesis.

Various codesign patterns have emerged to fuse the naturalness of neural language models with the precision and completeness of formal methods. One seeks to filter the outputs of a generative language model to satisfy a formal specification, typically by some form of rejection sampling. Alternatively, some attempt to steer language models to search for valid programs via a reinforcement learning or hybrid neurosymbolic approach. However, these approaches can introduce significant latency into the repair process and lack strong statistical guarantees.

Our work takes a more pragmatic tack - by incorporating the distance metric into a formal language, we attempt to exhaustively enumerate repairs by increasing distance, then use the language model to sort the resulting solutions by naturalness. The more constraints we can incorporate into formal language, the more efficient sampling becomes, and the more precise control we have over the output. This reduces the need for training a large, expensive language model to relearn syntax, and frees up our limited compute budget for more efficient search and ranking at inference time.

There is a delicate balance in formal methods between soundness and completeness. Often these two seem at odds because the target language is too expressive to achieve them both simultaneously. In syntax repair, we also care about naturalness. Fortunately, syntax repair is tractable enough to achieve all three by modeling the problem using language intersection. Completeness helps us to avoid missing simple repairs that might be easily overlooked, soundness guarantees all repairs will be valid, and naturalness ensures the most probable repairs receive the highest priority.

We have implemented our approach and demonstrated its viability as a tool for syntax assistance in real-world programming languages. Tidyparse [6] is capable of generating repairs for invalid source code in a range of programming languages. Unlike prior work, it is designed to work in the real-time interactive setting and generates a small list of repair suggestions that contain with high probability the author's intended repair. We plan to continue expanding the prototype's autocorrection functionality to cover an even broader family of real-world programming languages, in addition to Python. We envision a few primary use cases for it: (1) helping novice programmers become more quickly familiar with a new programming language, (2) autocorrecting common typos among proficient but forgetful programmers, (3) as a prototyping tool for PL designers and educators, and (4) as a pluggable library or service for IDEs, parsers and language servers.

[6]An artifact for Tidyparse is currently available as a browser application, supporting single-line syntax repairs in Python and other CFLs: https://tidyparse.github.io/.

## A IMPLEMENTATION

The implementation essentially consists of four stages, each dependent on its predecessor.

(1) `lev_build` : $\Sigma^n \times \mathbb{N} \rightarrow$ NFA – constructs a Levenshtein NFA from the broken string.
(2) `cfl_fixpt` : NFA $\times$ CFG $\rightarrow \mathbb{B}^{|Q|\times|Q|\times|V|}$ – computes the matrix exponential.
(3) `reg_build` : $\mathbb{B}^{|Q|\times|Q|\times|V|} \times$ CFG $\rightarrow$ RE – constructs the regular expression for $G_\cap$.
(4) `reg_dcode` : RE $\times$ WDFA $\times \mathbb{N} \rightarrow (\Sigma^+)^{k\approx 10^4}$ – returns a short list of high-probability repairs.

We will now explore the imperative pseudocode for each stage, starting with the Levenshtein automaton constructor, which is a straightforward translation of the inference rules in § 4.3.

**Algorithm 1** `lev_build` pseudocode

1: **procedure** `lev_build`$(\sigma : \Sigma^n, d_{\max} : \mathbb{N})$ ▷ Takes a string and maximum edit distance.
2: $\quad Q \leftarrow \varnothing, \delta \leftarrow \varnothing$
3: $\quad$ **for** $\langle h, j, i, k\rangle$ **in** $[0, n]^2 \times [0, d_{\max}]^2$ **do**
4:

$$\delta \leftarrow \delta \cup \left\{\begin{array}{llll} q_{h,i} \overset{[\neq\sigma_{j+1}]}{\longrightarrow} q_{j,k} & \text{if } h = j & \wedge\, i = k-1 \\ q_{h,i} \overset{[\neq\sigma_{j}]}{\longrightarrow} q_{j,k} & \text{if } h = j-1 & \wedge\, i = k-1 \\ q_{h,i} \overset{[=\sigma_{j}]}{\longrightarrow} q_{j,k} & \text{if } h = j-1 & \wedge\, i = k \\ q_{h,i} \overset{[=\sigma_{j}]}{\longrightarrow} q_{j,k} & \text{if } 1 \leq j-h-1 \leq d_{\max} & \wedge\, 1 \leq k-i \leq d_{\max} \end{array}\right\}$$

5: $\qquad Q \leftarrow Q \cup \{q_{h,i}, q_{j,k}\}$
6: $\quad I \leftarrow \{q_{0,0}\}, F \leftarrow \{q_{i,j} \mid n - i + j \leq d_{\max}\}$
7: $\quad$ **return** $\langle Q, \Sigma, \delta \subseteq Q \times (\Sigma \rightarrow \mathbb{B}) \times Q, q_\alpha, F\rangle$ ▷ Returns a [symbolic] Levenshtein automaton.

Next, the chart parser expects an acyclic NFA, a CNF grammar and returns a Boolean 3-tensor.

**Algorithm 2** `cfl_fixpt` pseudocode

**Require:** CFG must be in CNF and the NFA must be $\varepsilon$-free and acyclic (i.e., denote a finite language).
1: **procedure** `cfl_fixpt`$\big(\langle \Sigma, V, P, S\rangle : \text{CFG}, \langle Q, \Sigma, \delta, q_\alpha, F\rangle = A_\varnothing : \text{NFA}\big)$
2: $\quad R : \mathbb{B}^{|Q|\times|Q|} \leftarrow \left[\blacksquare \textbf{ if } \exists \sigma \in \Sigma^+ \mid q \overset{\sigma}{\rightsquigarrow} q' \textbf{ else } \square\right]_{q,q':Q}$ ▷ Solve for reachability matrix.
3: $\quad M : \mathbb{B}^{|Q|\times|Q|\times|V|} \leftarrow \left[\blacksquare \textbf{ if } \exists s : \Sigma \mid (v \rightarrow s) \in P \wedge (q \overset{\varphi}{\rightarrow} q') \in \delta \wedge \varphi(s) \textbf{ else } \square\right]_{q,q':Q,\,v:V}$
4: $\quad$ **for** $i$ **in** $\big[0, |A_\varnothing|\big]$ **do** ▷ Solves matrix exponential, $M_\infty = \exp(M_0)$.
5: $\qquad$ DONE $\leftarrow \blacksquare$
6: $\qquad$ **for** $\langle p, r, w\rangle$ **in** $Q^2 \times V$ **do** ▷ Iterates one step of $M_{i+1} = M_i + M_i^2$.
7: $\qquad\quad$ **if** $M[p, r, w]$ **or not** $R[p, r]$ **then continue**
8: $\qquad\quad Q_{pr} \leftarrow \big\{q : Q \mid R[p, q] \wedge R[q, r]\big\}$ ▷ Consider reachable states between p and r.
9: $\qquad\quad M[p, r, w] \leftarrow \blacksquare$ **if** $\exists q : Q_{pr}, x, z : V \mid M[p, q, x] \wedge M[q, r, z] \wedge (w \rightarrow xz) \in P$ **else** $\square$
10: $\qquad\quad$ **if** $M[p, r, w]$ **then** DONE $\leftarrow \square$
11: $\qquad$ **if** DONE **then break**
12: $\quad$ **return** $M$ ▷ Returns the completed Boolean parse chart.

Note we may short-circuit for three reasons, if: $M_{i+1} = M_i$, when two states $q, q'$ are unreachable, or whenever a $\langle q, q', v\rangle$ is already present. Once we obtain $M_\infty$, we can immediately tell whether $\ell_\cap \neq \varnothing$ by checking whether $M_\infty[q_\alpha, q_\omega, S] = \blacksquare$ for some $q_\omega : F$. Otherwise, if no such $q_\omega$ exists, then $\ell_\cap$ must be empty and $d_{\max}$ should be enlarged before proceeding.

Now we can perform a second sweep over nonempty entries of the Boolean parse chart, reconstructing the provenance of each $\langle q, q', v\rangle$ constituent. For compactness it will be convenient to use a pointer-based representation of the regular expression instead of manipulating strings.

**Algorithm 3** `reg_build` pseudocode

**Require:** Same as `cfl_fixpt` (Alg. 2), $M_{\mathbb{B}}[q_\alpha, q_\omega, S] = \blacksquare$ for some $q_\omega : F$, and $M_{\mathbb{B}} = M_{\mathbb{B}} + M_{\mathbb{B}}^2$.

1: **procedure** `reg_build`$\big(M_{\mathbb{B}} : \mathbb{B}^{|Q|\times|Q|\times|V|}, \langle \Sigma, V, P, S\rangle : \text{CFG}, \langle Q, \Sigma, \delta, q_\alpha, F\rangle = A_\varnothing : \text{NFA}\big)$

2: $\quad P : \mathbb{B}^{|Q|\times|Q|} \leftarrow \big[\blacksquare \textbf{ if } \exists q : Q, v, v' : V \mid M_{\mathbb{B}}[p, q, v] \wedge M_{\mathbb{B}}[q, r, v'] \textbf{ else } \square\big]_{p,r:Q}$

3: $\quad M : \text{RE}^{|Q|\times|Q|\times|V|} \leftarrow \big[\{s : \Sigma \mid M[q, q', v] \wedge (q \xrightarrow{\varphi} q') \in \delta \wedge (v \to s) \in P \wedge \varphi(s)\}\big]_{q,q':Q,v:V}$

4: $\quad$ **for** $i$ **in** $\big[0, |A_\varnothing|\big]$ **do**

5: $\quad\quad M' \leftarrow M$

6: $\quad\quad$ **for** $\langle p, r, w\rangle$ **in** $Q^2 \times V$ **do**

7: $\quad\quad\quad$ **if not** $M_{\mathbb{B}}[p, r, w]$ **then continue**

8: $\quad\quad\quad Q_{pr} \leftarrow \big\{q : Q \mid P[p, q] \wedge P[q, r]\big\}$ ▷ Consider parseable states between p and r.

9: $\quad\quad\quad M'[p, r, w] \leftarrow M[p, r, w] \vee \bigvee_{\substack{q:Q_{pr}\\ x,z:V}} \big\{M[p, q, x] \cdot M[q, r, z] \mid (w \to xz) \in P\big\}$

10: $\quad\quad$ **if** $M = M'$ **then break else** $M \leftarrow M'$

11: $\quad$ **return** $\bigvee_{q_\omega : F} M[q_\alpha, q_\omega, S]$ ▷ Union regexes for all total parses yielding S.

Finally, once we have the expression for $G_\cap$, we can decode it to extract a small set of candidates. We use two priority queues to store partial and total trajectories, which are ranked by the WDFA language prior $\alpha_{\hat{\theta}}$ from § 6.4. During best-first search, each partial trajectory carries the current residual expression and WDFA state. A one-token extension by $a$ advances the WDFA to $\delta(q, a)$ and adds $\ln w(q, a)$ to the running score; when a trajectory terminates, we add the final weight $\ln w_F(q)$ before placing it on the total queue and the top-$k$ total trajectories are returned.

**Algorithm 4** `reg_dcode` pseudocode

**Require:** $\alpha_{\hat{\theta}} = \langle Q_\alpha, \Sigma, \delta, q_0, F_\alpha, w, w_F\rangle$, a positive-weight WDFA (§ 6.4), and $\ell_\cap = \mathcal{L}(e) \subseteq \mathcal{L}(\alpha_h)$.

1: **procedure** `reg_dcode`$\big(e : \text{RE}, \alpha_{\hat{\theta}} : \text{WDFA}, k : \mathbb{N}\big)$

2: $\quad \mathcal{T} \leftarrow [\,], \mathcal{E} \leftarrow \big[\langle \varepsilon, e, q_0, 0\rangle\big]$ ▷ Initialize total and partial trajectories.

3: $\quad$ **repeat**

4: $\quad\quad \langle \sigma, e, q, p\rangle \leftarrow$ **pop** $\mathcal{E}_0$ **off** $\mathcal{E}$ ▷ Prefix, residual, current state, partial probability.

5: $\quad\quad$ **if** $\varepsilon \in \mathcal{L}(e) \wedge q \in F_\alpha$ **then** $\mathcal{T} \leftarrow \mathcal{T} \mathbin{+\!\!+} \big[\langle \sigma, p + \ln w_F(q)\rangle\big]$

6: $\quad\quad \mathcal{E}' \leftarrow \big[\langle \sigma \cdot a, \partial_a e, \delta(q, a), p + \ln w(q, a)\rangle \mid a \in \texttt{next}(e)\big]$

7: $\quad\quad \mathcal{E} \leftarrow \big[\langle \sigma, e, q, p\rangle \in (\mathcal{E} \mathbin{+\!\!+} \mathcal{E}')$ **sorted by decreasing** $p\big]$

8: $\quad$ **until** interrupted or $\mathcal{E}$ is empty.

9: $\quad$ **return** $[\sigma \mid \langle \sigma, p\rangle \in (\mathcal{T}$ **sorted by decreasing** $p)_{0..k}]$ ▷ Skim off top-$k$ repairs by $P_{\hat{\theta}}(\sigma)$.

As shown in Alg. 4, we decode the regular expression $e$, representing exactly $\ell_\theta$, over which the WDFA $\alpha_{\hat{\theta}}$ soundly overapproximates and orders exploration. This enables us to quickly generate a representative subset of $\ell_\cap$ under the finite-state prior $P_{\hat{\theta}}$ within a fixed latency budget or else terminate early should we succeed in exhaustively generating it. After reranking, truncation, and cosmetic postprocessing, the resulting repairs may then be displayed to the user.

## A.1 GPU translation

The foregoing architecture can be translated to a series of high-performance GPU kernels. Our strategy will be to maximize GPU utilization by distributing the workload for each stage across as many independent threads as we can simultaneously dispatch. Each thread will be responsible for writing to a dedicated portion of a shared buffer without locking or external communication.

We will make the simplifying assumption that each GPU kernel is a pure function that takes as input a coordinate triple $r, c, v : \mathbb{N}$ and one or more flat buffers $b_1 : \mathbb{N}^{d_1}, \ldots b_n : \mathbb{N}^{d_n}$ containing encoded data, does some arithmetic, and returns a single output buffer, $b_{\text{out}} : \mathbb{N}^d$. On a GPU, all memory must be sized ahead of time, as dynamic allocation is forbidden during a GPU kernel's execution. The main challenge of GPU programming then becomes careful memory management and efficiently mapping aggregate datatypes to and from the integers. Conceptually, each $\langle r, c, v\rangle$ triple will be dispatched to a single GPU thread with global read access to the input buffers and exclusive write access to a contiguous region of the output buffer. Consistent with Theorem 3.1, each thread will correspond to a single processor, with $|Q|^2|V|$ threads in total. Absent a GPU, this can be rewritten as a triply-nested loop, subject to additional latency.

For the CFG and NFA datatypes, we elect to use a dense representation $\mathbb{B}^{|V|\times|V|\times|V|}$ and $\mathbb{B}^{|Q|\times|Q|\times|\Sigma|}$ due to the tripartite coordinate structure and thread dispatching API. While these datatypes can be encoded sparsely as $\mathbb{N}^{3|P|}$ and $\mathbb{N}^{3|\delta|}$, for most repair instances and memory configurations representation size is not a bottleneck. It will be helpful to define characteristic functions `nt_enc` $: \Sigma \rightarrow 2^V$, `nt_dec` $: V \rightarrow 2^\Sigma$ for nonterminal encoding and decoding, and index sets $\Sigma \hookrightarrow \mathbb{N}$, $V \hookrightarrow \mathbb{N}$, $Q \hookrightarrow \mathbb{N}$ for getting in and out of the `uint` domain, with $|V|, |Q| \lesssim 10^3$ adjustable upward if memory permits.

The parse chart $M$ can be represented as a bit-packed integer matrix $\texttt{uint32}^{|Q|\times|Q|\times|V|}$, whose layout testifies to four properties of each $\langle q, q', v\rangle$ triple: (1) the first bit encodes the dis/equality predicate $\varphi$, (2) the next 25 bits designate terminal participation $\big(\text{if } \exists s : \Sigma.\varphi(s) \land (q \overset{\varphi}{\rightarrow} q') \in \delta\big)$, (3) the next five bits memoize the minimum $i_{\min}$ such that $M_{i_{\min}}[q, q', v]\&1 = \blacksquare$ for short-circuiting (see Line #7 of Alg. 3), and (4) the lowest-order bit denotes parsability, i.e., $q \rightsquigarrow q' \vdash v$. Note the decoder must acknowledge the possibility that $v$ can simultaneously parse (a) an arc $q \rightarrow q'$ and (b) a path $q \rightsquigarrow q'$, so each branch can be explored. This is depicted below in little-endian format:

$$\Big[\overset{\overset{=/\neq}{\Updownarrow}}{\blacksquare}, \overbrace{\square,\square,\ldots,\blacksquare,\square,\blacksquare}^{s\,:\,2^{\Sigma}\hookrightarrow\mathbb{B}^{25}}, \overbrace{\square,\square,\square,\square,\blacksquare}^{i_{\min}\,:\,\mathbb{N}_{\leq 32}\leftrightarrow\mathbb{B}^{5}}, \overset{\overset{\exists\, v:V.\,q\rightsquigarrow q'\vdash v}{\Updownarrow}}{\blacksquare}\Big] : \texttt{uint32}$$

Once `cfl_fixpt` (Alg. 2) is complete, we can calculate the total amount of memory needed to allocate $G_\cap$ by counting constituents in the parse chart. Being an algebraic datatype, the RE can be flattened according to a variety of allocation models. We will use the following memory layout,

$$\Big[\overbrace{\underset{\texttt{bp}_0}{2},\underset{\texttt{bp}_1}{7},\ldots,\underset{\texttt{bp}_{c-2}}{1},\underset{\texttt{bp}_{c-1}}{3}}^{\texttt{bp_counts}}, \overbrace{0,4,\ldots,\underset{\texttt{bp}_{c-2}}{n-8},\underset{\texttt{bp}_{c-1}}{n-6}}^{\texttt{bp_offsets}}, \overbrace{\underbrace{\underline{59,83},\underline{64,152}}_{\texttt{bp}_0},\ldots,\underbrace{\underline{34,83}}_{\texttt{bp}_{c-2}},\underbrace{\underline{22,74},\underline{74,90},\underline{16,66}}_{\texttt{bp}_{c-1}}}^{\texttt{bp_storage}}\Big] : \texttt{uint32}^n$$

where each $\texttt{bp}_i$ represents a nonempty $\langle q, q', v\rangle$ constituent with at least one back-pointer pair, `bp_count`$(p, r, w) = \big|\{\langle q, x, z\rangle \mid M[p, q, x] \land M[q, r, z] \land (w \rightarrow xz) \in P\}\big|$ counts the number of unique backpointers held by each nonterminal $w$ parseable from $p \rightsquigarrow r$, and `bp_storage` stores pointers to memory locations in the same data structure. These pointers should also be tied to locations in the parse chart $M[q, q', v]$ to recover the terminal subsets for terminal productions.

In total, the GPU should have at least 4 GB of onboard memory to accommodate language intersections with up to $10^3$ states and nonterminals $\big(|Q|^2 \times |V| \lesssim 10^6 \times 10^3 \times 32 \text{ bits} \approx 4 \text{ GBs}\big)$, however occupancy can be roughly halved by exploiting the upper-triangular structure of $M$.

## A.2 Training the reranker

After decoding, we have a list of repair candidates that are all valid, nearby and at least somewhat plausible, however it is possible this list may be quite long. No reasonable user would be expected to skim through more than a few dozen candidates to select their intended repair, especially since they could have presumably written it themselves in a few seconds. So, we will proceed to rerank the list. If we can guarantee the candidate repairs are sufficiently exhaustive, they should include with high probability the true repair, which then need only be surfaced into the user's field of view.

The ensuing method falls under the umbrella of the *learning-to-rank* (LTR) problem in machine learning – using their terminology, the broken code snippet would be called a *query* and the list of repairs, *documents*. To discuss the reranker, we must now overload some concepts, so the reader is trusted to contextually interpret $\mathcal{L}$ as denoting a *loss* instead of a language, and the derivative [7] as the directional rate of change of a differentiable manifold over the parameter space of a neural network. Matrix multiplication remains more or less the same, except now over the reals.

The reranker employs a transformer-encoder architecture to map both the query (broken code snippet, denoted $\underset{\sim}{\sigma} : \bar{\ell}$) and the document (candidate repair, denoted $\sigma : \ell_\cap$) to a $p$-dimensional real vector space, $\mathbb{R}^p$. We elide the definition of a transformer-encoder (see Strobl et al.'s survey [58]), except to say that it is a function, $\mathrm{E}_\theta$, which takes a string and a positional encoding, and returns an embedding, $\mathrm{E}_\theta : (\Sigma \times \mathbb{N})^n \to \mathbb{R}^p$ where $\theta$ are learnable parameters. We introduce a second encoder, $E'_\theta : (\Sigma \times \mathbb{N} \times \mathbb{Z}_4)^{[m,n]} \to \mathbb{R}^p$, which takes in a Levenshtein alignment $(\mathrm{LA} : \Sigma^n \times \Sigma^m \to \mathbb{Z}_4^{[m,n]})$ tracking edit locations and types for each query-document pair. Finally, a multilayer perceptron $(\mathrm{MLP}_\theta : \mathbb{R}^p \times \mathbb{R}^p \to \mathbb{R}^+)$ processes the embedding to produce a scalar relevance score:

$$f_\theta(\underset{\sim}{\sigma}, \sigma) : \Sigma^n \times \Sigma^m \to \mathbb{R}^+ = \mathrm{MLP}_\theta\Big(\mathrm{E}_\theta\big(\underset{\sim}{\sigma}, [i]_{i\in[0,n)}\big), \mathrm{E}'_\theta\big(\sigma, [i]_{i\in[0,m)}, \mathrm{LA}(\underset{\sim}{\sigma}, \sigma)\big)\Big) \tag{11}$$

Our training objective will be to minimize the tempered *softmax* or listwise cross-entropy loss,

$$\mathcal{L}(\theta) = -\sum_{q\in\mathcal{Q}} \log\left(\frac{\exp\big(f_\theta(q, d^*)\tau^{-1}\big)}{\sum_{d\in\mathcal{D}_q} \exp\big(f_\theta(q, d)\tau^{-1}\big)}\right) \tag{12}$$

where $\mathcal{Q}$ is the set of training queries, $\mathcal{D}_q$ is the set of candidate repairs for query $q$, and $d^* \in \mathcal{D}_q$ is true repair. The temperature parameter, $\tau$ controls the sharpness of the softmax distribution, encouraging parameter settings that result in the true repair being assigned higher priority – the closer to zero, the greater the loss will be for underestimating the relevance of the true repair.

More concretely, we depict a single instance of the training data in Fig. 19. The reranking model sees a (1) query, (2) document, (3) positional encoding and (4) Levenshtein alignment, and returns a numerical score. Once the relevance scores are obtained, we calculate the cross-entropy loss across the top-$k$ scoring documents and backpropagate. In practice, this update is averaged across a batch $\langle q_i, [d_j]_{0\ldots k}\rangle_{i=0\ldots|B|\approx 16}$ of repair instances to reduce noise. The batch update rule is a standard variant of stochastic gradient descent $\big(\theta' \leftarrow \theta - \alpha\nabla_\theta\mathcal{L}(\theta)\big)$ with momentum (AdamW), where the learning rate is in the range $\alpha \approx 10^{-4}$. A more exhaustive description of the architectural details and hyperparameter settings used for training the reranker can be found in Appendix E.

| $q$: | NAME | = | NAME | ) | NAME | | |
|---|---|---|---|---|---|---|---|
| PE: | 0 | 1 | 2 | 3 | 4 | | |
| $d_1$: | NAME | = | NAME | ( | NAME | ) | |
| PE: | 0 | 1 | 2 | 3 | 4 | 5 | |
| LA: | 0 | 0 | 0 | 2 | 0 | 1 | |
| $d_2$: | NAME | ( | NAME | ) | NAME | | |
| PE: | 0 | 1 | 2 | 3 | 4 | | |
| LA: | 0 | 2 | 0 | 0 | 3 | | |
| $d_3$: | NAME | = | NAME | . | NAME | ( | ) |
| PE: | 0 | 1 | 2 | 3 | 4 | 5 | 6 |
| LA: | 0 | 0 | 0 | 2 | 0 | 1 | 1 |

Fig. 19. Transformer data encoding.

[7] There is a connection to Brzozowski's derivative, but to refrain from digression here we refer the reader to [24] for details.

## B LEVENSHTEIN TOPOLOGY AND MATRICES

These are useful for visually checking different implementations.

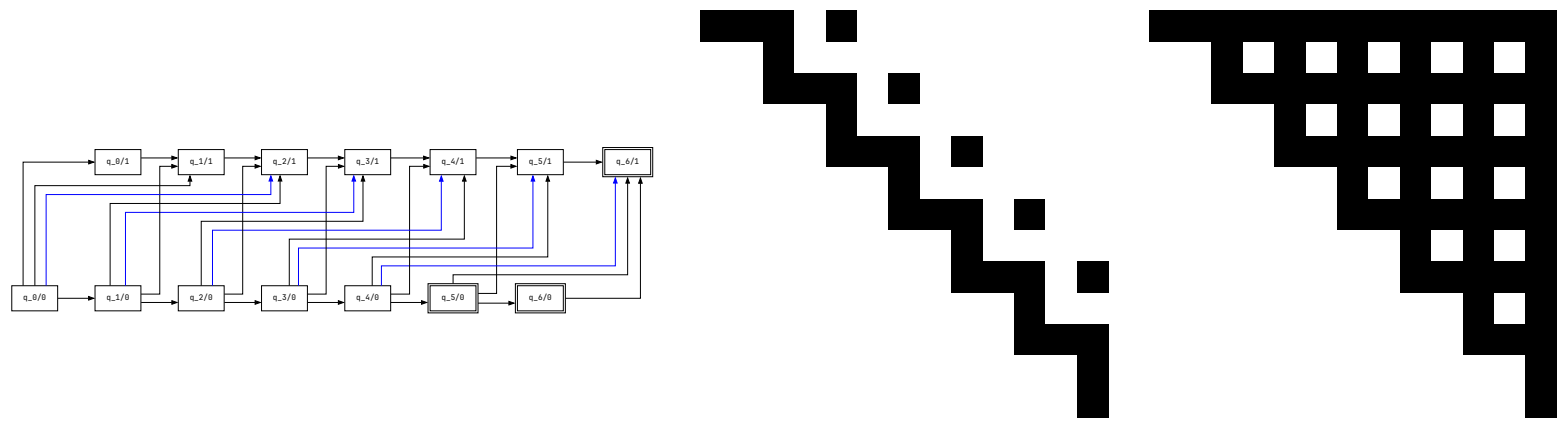

Fig. 20. Lev($|\sigma|$=6, $\Delta$=1) automaton, adjacency and reachability matrix.

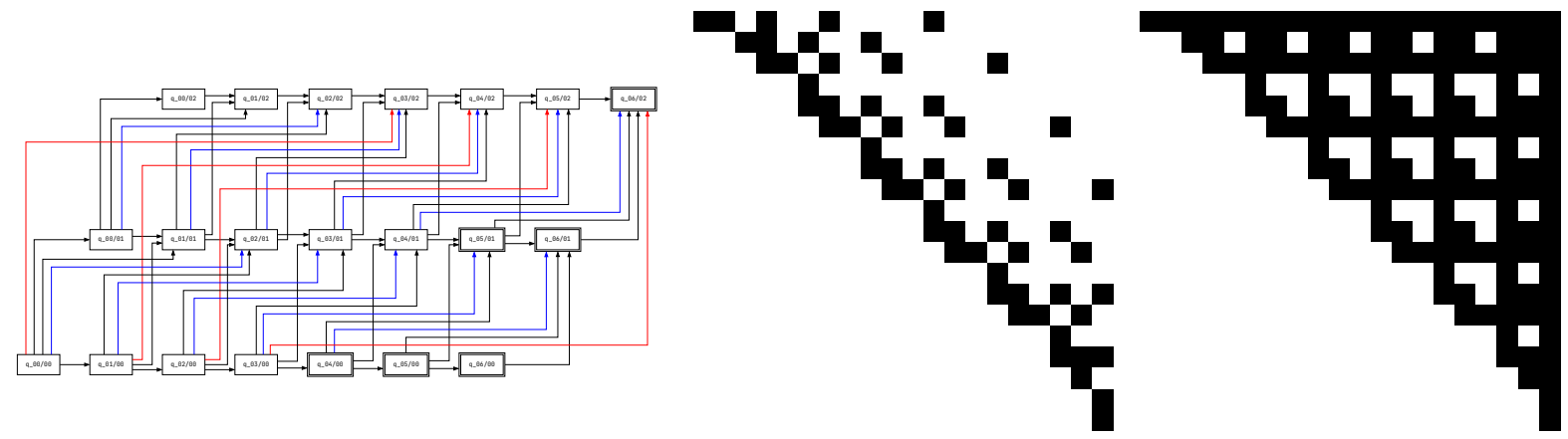

Fig. 21. Lev($|\sigma|$=6, $\Delta$=2) automaton, adjacency and reachability matrix.

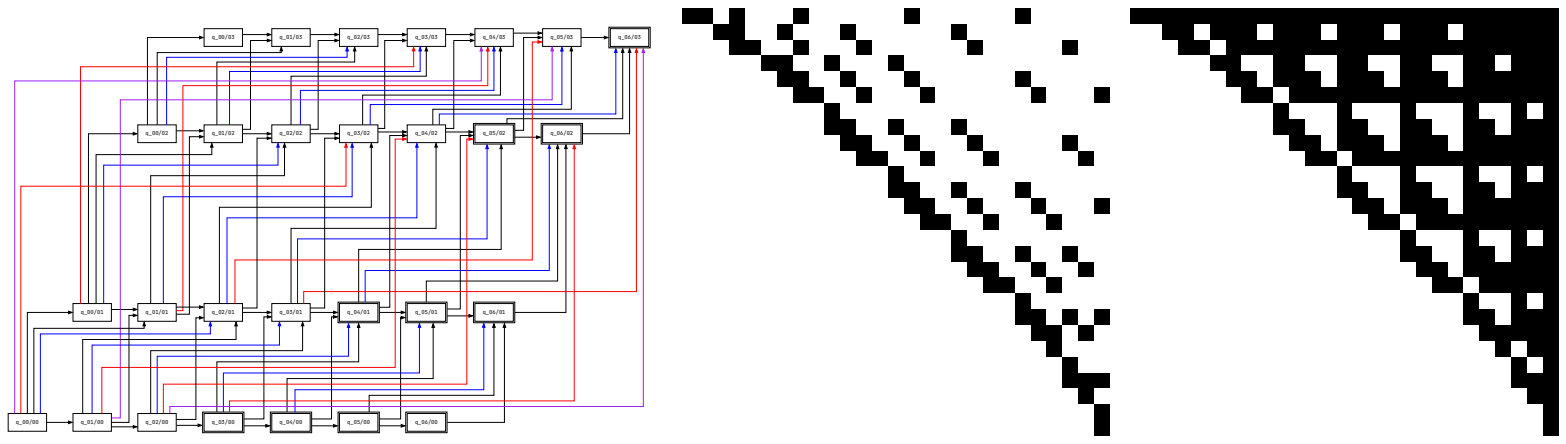

Fig. 22. Lev($|\sigma|$=6, $\Delta$=3) automaton, adjacency and reachability matrix.

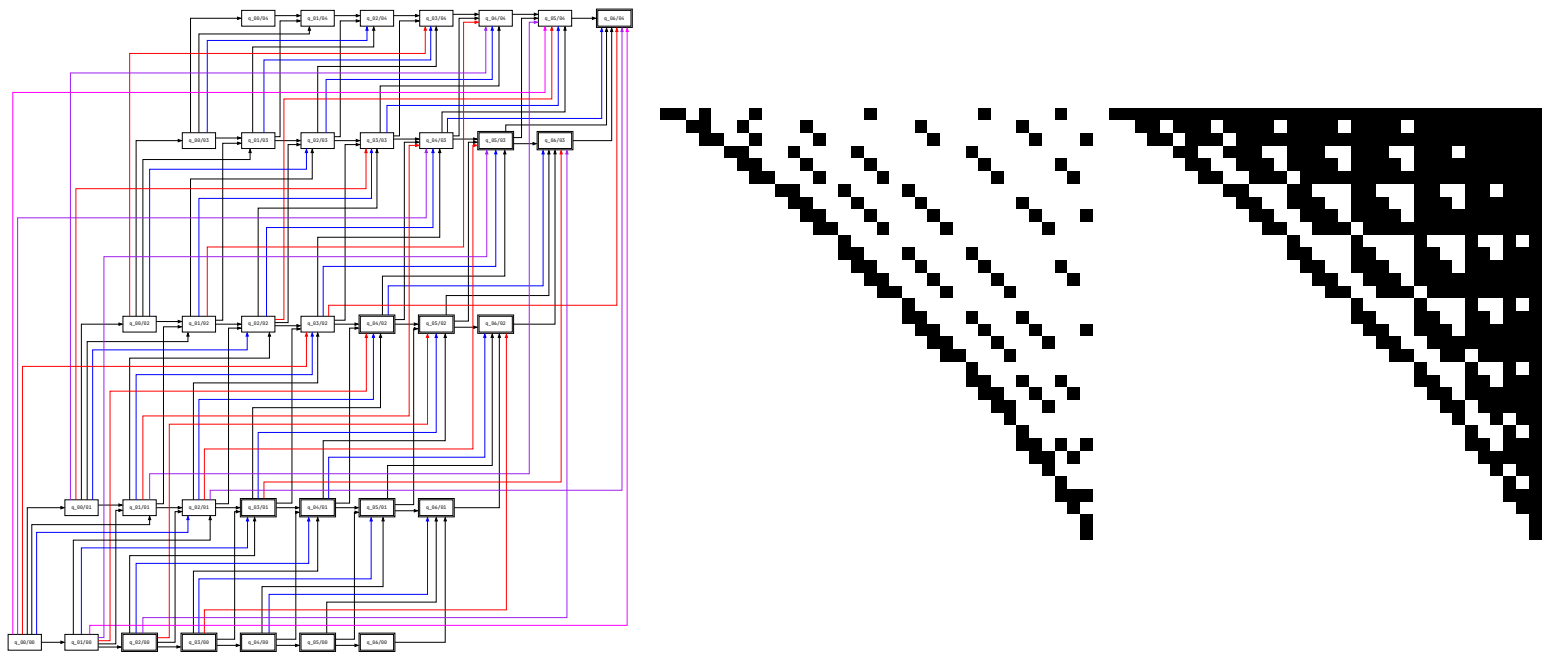

Fig. 23. Lev($|\sigma|$=6, $\Delta$=4) automaton, adjacency and reachability matrix.

## C EXAMPLE REPAIRS

Below, we provide a few representative examples of broken code snippets and the corresponding human repairs that were successfully ranked first by our method. On the left is a complete snippet fed to the model, and on the right, the corresponding human repair that was correctly predicted.

<table>
<tr><th>Original broken code</th><th>First predicted repair</th></tr>
<tr><td><pre><code>form sympy import *
x = Symbol('x', real=True)
x, re(x), im(x)</code></pre></td><td><pre><code>from sympy import *
x = Symbol('x', real=True)
x, re(x), im(x)</code></pre></td></tr>
<tr><td><pre><code>result = yeald From(item.create())
raise Return(result)</code></pre></td><td><pre><code>result = yield From(item.create())
raise Return(result)</code></pre></td></tr>
<tr><td><pre><code>df.apply(lambda row: list(set(row['ids'))))</code></pre></td><td><pre><code>df.apply(lambda row: list(set(row['ids'])))</code></pre></td></tr>
<tr><td><pre><code>sum(len(v) for v items.values()))</code></pre></td><td><pre><code>sum(len(v) for v in items.values())</code></pre></td></tr>
<tr><td><pre><code>def average(values):
  if values == (1,2,3):
    return (1+2+3)/3
  else if values == (-3,2,8,-1):
    return (-3+2+8-1)/4</code></pre></td><td><pre><code>def average(values):
  if values == (1,2,3):
    return (1+2+3)/3
  elif values == (-3,2,8,-1):
    return (-3+2+8-1)/4</code></pre></td></tr>
<tr><td><pre><code>dict = {
  "Jan": 1
  "January": 1
  "Feb": 2 # and so on
}</code></pre></td><td><pre><code>dict = {
  "Jan": 1,
  "January": 1,
  "Feb": 2 # and so on
}</code></pre></td></tr>
<tr><td><pre><code>class MixIn(object)
  def m():
    pass

class classA(MixIn):

class classB(MixIn):</code></pre></td><td><pre><code>class MixIn(object):
  def m():
    pass

class classA(MixIn): pass

class classB(MixIn): pass</code></pre></td></tr>
</table>

## D RAW DATA

Raw data from Precision@k experiments across snippet length and Levenshtein distance from § 7.3. $|\sigma|$ is the snippet length and $\Delta$ is the Levenshtein distance between the broken code and human fix in terms of lexical tokens. For Precision@1, we report the percentage of top-1 repairs matching the human repair. For Precision@All, we report the percentage of true repairs contained within the top 20,000 samples. Each entry in the following table represents a pairwise disjoint subset of $D_{\text{test}}$, with at least 50 distinct Python syntax errors and repairs matching the length and distance criteria, sampled uniformly from the full Stack Overflow dataset [64].

| | $\Delta$ | Precision@1 | | | | | | | |
|---|---|---|---|---|---|---|---|---|---|
| $\|\sigma\|$ | | $(0, 10)$ | $[10, 20)$ | $[20, 30)$ | $[30, 40)$ | $[40, 50)$ | $[50, 60)$ | $[60, 70)$ | $[70, 80)$ |
| Tidyparse | 1 | 0.578 | 0.459 | 0.421 | 0.416 | 0.330 | 0.269 | 0.282 | 0.313 |
| | 2 | 0.387 | 0.376 | 0.428 | 0.456 | 0.475 | 0.346 | 0.460 | 0.380 |
| | 3 | 0.241 | 0.129 | 0.120 | 0.105 | 0.120 | 0.185 | 0.218 | 0.179 |
| Seq2Parse | 1 | 0.35 | 0.41 | 0.40 | 0.37 | 0.31 | 0.29 | 0.27 | 0.21 |
| | 2 | 0.12 | 0.13 | 0.14 | 0.12 | 0.11 | 0.11 | 0.10 | 0.12 |
| | 3 | 0.03 | 0.07 | 0.08 | 0.09 | 0.09 | 0.02 | 0.07 | 0.06 |
| BIFI | 1 | 0.20 | 0.33 | 0.32 | 0.27 | 0.21 | 0.21 | 0.25 | 0.18 |
| | 2 | 0.18 | 0.18 | 0.21 | 0.19 | 0.19 | 0.18 | 0.11 | 0.11 |
| | 3 | 0.02 | 0.02 | 0.03 | 0.02 | 0.03 | 0.05 | 0.03 | 0.02 |
| GPT-5.4 | 1 | 0.485 | 0.401 | 0.477 | 0.459 | 0.453 | 0.472 | 0.629 | 0.553 |
| | 2 | 0.040 | 0.094 | 0.062 | 0.077 | 0.056 | 0.075 | 0.222 | 0.158 |
| | 3 | 0.089 | 0.063 | 0.052 | 0.050 | 0.043 | 0.100 | 0.054 | 0.029 |
| | | Precision@All | | | | | | | |
| Tidyparse | 1 | 1.000 | 1.000 | 1.000 | 1.000 | 1.000 | 1.000 | 1.000 | 1.000 |
| | 2 | 1.000 | 1.000 | 1.000 | 1.000 | 1.000 | 1.000 | 1.000 | 1.000 |
| | 3 | 0.431 | 0.323 | 0.333 | 0.360 | 0.413 | 0.459 | 0.490 | 0.500 |
| BIFI | 1 | 0.65 | 0.67 | 0.70 | 0.65 | 0.60 | 0.62 | 0.60 | 0.64 |
| | 2 | 0.52 | 0.41 | 0.37 | 0.32 | 0.27 | 0.27 | 0.21 | 0.24 |
| | 3 | 0.20 | 0.13 | 0.08 | 0.17 | 0.15 | 0.18 | 0.17 | 0.07 |

## E HYPERPARAMETER SETTINGS

Below is a listing of the hyperparameter settings used for training the reranking model:

- Input dimension: 100
- Encoder dimension: 512
- Attention heads: 4
- Encoder layers: 4
- Vocab size, $|\Sigma| = 94$
- Learning rate, $\alpha = 10^{-3}$
- Temperature, $\tau = 10^{-1}$
- Optimizer: AdamW
- Negative rate: $10^{-2}$
- Batch size: 8
- Dropout: $10^{-1}$
- Activation: GELU

The full parameters $\theta$ are partitioned into two sets, $\theta_e, \theta_r$, for the encoder and reranker layers. The encoder is pretrained on next-token prediction, then fine-tuned on the reranking task. During optimization, we use a smaller learning rate ($\alpha = 10^{-5}$) so as not to disturb the pretrained encoder parameters and a larger learning rate ($\alpha = 10^{-4}$) for the reranker parameters. We train the encoder for $2 \times 10^4$ steps and the reranker for $1.3 \times 10^4$ steps, taking $\sim 4$ hours on an Nvidia H100 GPU.

### E.1 WDFA training

For WDFA-guided retrieval, we construct the $h = 2$ Nederhof approximation for the Python grammar, remove $\varepsilon$-arcs and determinize it. The resulting serialized WDFA has 2,115 states and 82,257 transitions. We then fit a locally normalized stochastic model to the deterministic support using a training corpus consists of 6,097,537 Python snippets and 243,521,023 total repair tokens. For each snippet, we increment every traversed arc count. This estimates

$$w(q,a) = \log \frac{c(q,a)+\lambda}{D_q} \qquad D_q = \sum_{b\in \mathrm{out}(q)} c(q,b) + \lambda\big(\lvert\mathrm{out}(q)\rvert + \mathbb{1}[q \in F]\big).$$

We use Lidstone smoothing with $\lambda = 1$.

## F SYMBOLS AT A GLANCE

Below we provide an inexhaustive listing of some common notation used throughout this paper.

| Notation | Meaning |
|---|---|
| $G = \langle \Sigma, V, P, S \rangle$ | CFG with terminals, $\Sigma$, nonterminals, $V$, productions, $P$, and start symbol, $S$. |
| $A = \langle Q, \Sigma, \delta, q_\alpha, F \rangle$ | Automaton with states, $Q$, transitions, $\delta$, start state, $q_\alpha$ and final states, $F$. |
| $\underset{\sim}{\sigma}, \sigma^\star$ | Invalid string with known target language, and ground-truth repair. |
| $A_\varnothing, \lvert A_\varnothing \rvert$ | Acyclic automaton and its corresponding span (see Thm. 3.1). |
| $G'$ | Chomsky Normal Form (CNF) grammar. |
| $G_\cap, e_\cap$ | Grammar and regular expression formed by the CFG-NFA intersection. |
| $\ell_\cap, \mathcal{L}(G_\cap)$ | Intersection language generated by some $G_\cap$. |
| $L(\underset{\sim}{\sigma}, k)$ | Levenshtein automaton of radius $k$ for a broken string, $\underset{\sim}{\sigma}$. |
| $[\ldots]$ | Related to the symbolic predicate in the Levenshtein automaton. |
| $d_{\max}$ | Maximum permitted Levenshtein edit distance (repair radius). |
| $M$ | Matrix encoding the product construction $\mathcal{L}(G) \cap \mathcal{L}\big(L(\underset{\sim}{\sigma}, k)\big)$. |
| $P@k$ | Precision at rank $k$ evaluation metric. |

## G GPT-5.4 REPAIR PROMPT

For the GPT-5.4 baseline in Fig. 14, we queried the OpenAI Responses API with model `gpt-5.4-mini`, `max_output_tokens=1024`, `store=false`, no temperature override, and no reasoning parameter. Each request used a single user message containing the following prompt, with `$brokenTokens` replaced by the tokenized broken snippet:

```
You are repairing a Python syntax error in tokenized form.

The input and output are whitespace-separated lexer tokens, not source code.
Identifiers and literals are abstracted as NAME, NUMBER, and STRING.
Structural tokens such as NEWLINE, INDENT, DEDENT, and ENDMARKER must remain tokens.

Return exactly one repaired token sequence in the same whitespace-separated token format.
Return the complete repaired sequence, including all unchanged tokens from the input.
Do not return only the changed token, only the edit, or a patch.
Make the smallest syntactic repair that is most likely to be the author's intended code.
Do not return markdown, XML tags, explanations, alternatives, or source code.

Broken token sequence:
$brokenTokens
```

The response was interpreted as a single whitespace-separated token sequence after stripping code fences, optional `Repair:` prefixes, surrounding quotes, and redundant whitespace. Across 2,238 repair instances in the test set, this cost roughly US $1 to evaluate as of June 2026 pricing.